\documentclass[12pt, a4paper]{article}
\usepackage{lineno}
\usepackage{color}
\usepackage{graphicx}
\usepackage[toc,page]{appendix}
\usepackage{caption}
\usepackage{amsmath}
\usepackage{subcaption}
\usepackage{hyperref}
\usepackage{mathtools}
\usepackage{float}
\usepackage{rotating}
\usepackage{url}
\usepackage{enumitem}
\usepackage[table]{xcolor}
\usepackage{multirow}
\usepackage[font=scriptsize,labelfont=bf,skip=6pt]{caption}
\usepackage[T1]{fontenc}
\usepackage{authblk}
\font\myfont=cmr12 at 15pt

\begin{document}
\title{\myfont{Layout and Assembly Technique of the GEM Chambers for the Upgrade of the CMS First Muon Endcap Station}}

\author[27]{D. Abbaneo}
\author[27]{D. Abbas}
\author[10]{M. Abbrescia}
\author[16]{A. Ahmad}
\author[8]{A. Ahmed}
\author[16]{W. Ahmed}
\author[28]{C.Ali}
\author[16]{I. Asghar}
\author[27]{P. Aspell}
\author[5]{Y. Assran}

% ...
\affil[5]{Academy of Scientific Research and Technology - ENHEP, Cairo, Egypt}
% ...
\affil[8]{Delhi University, Delhi, India}
% ...
\affil[10]{ Politecnico di Bari, Universit`a di Bari and INFN Sezione di Bari, Bari, Italy}
% ...
\affil[16] {National Center for Physics, Islamabad, Pakistan}
% ...
\affil[27]{CERN, Geneva, Switzerland}
\affil[28]{Texas $\&$ University - Qatar (associated with Texas A$\&$M University, USA), Doha, Qatar}
% ...

\maketitle
\begin{abstract}

Triple-GEM detector technology was recently selected by CMS for a part of the upgrade of its forward muon detector system as GEM detectors provide a stable operation in the high radiation  environment expected during the future High-Luminosity phase of the Large Hadron Collider (HL-LHC). In a first step, GEM chambers (detectors) will be installed  in the innermost  muon endcap station in the    $1.6<\left|\eta\right|<2.2$ pseudo-rapidity region, mainly to control level-1 muon trigger rates after the second LHC Long Shutdown.  These new chambers  will add redundancy to the muon system in the $\eta$-region where the background rates are high, and the bending of the muon trajectories due to the CMS magnetic field is small. A novel construction technique for such chambers has been developed in such a way where foils are mounted onto a single stack and then uniformly  stretched mechanically, avoiding the use of spacers and glue inside the active gas volume. We describe the layout, the stretching mechanism and the overall
assembly technique of such GEM chambers. 

\end{abstract}
\begin{keywords}
CMS, GEM, High Luminosity LHC
\end{keywords}
\clearpage 

\tableofcontents

\clearpage 
\section{Introduction}

The current schedule of the CERN Large Hadron Collider (LHC) includes several long shutdown periods to allow for upgrades of the  accelerator complex as well as the experiments.   The ultimate upgrade of the LHC to enable operation at a proton-proton center-of-mass energy of 14~TeV with an instantaneous luminosity gradually increasing up to 5-7 $\times$ 10$^{34}$ cm$^{-2}$s$^{-1}$ is referred to as the High Luminosity LHC (HL-LHC). After the two-year long second Long Shutdown (LS2) starting in 2019, the  instantaneous luminosity will exceed 2 $\times$ 10$^{34}$ cm$^{-2}$ s$^{-1}$, while the third  Long Shutdown (LS3) scheduled for 2024-2026 will bring the LHC luminosity to its ultimate level of about 5 to 7 times its design value.  The HL-LHC (or Phase-2) period starting after LS3 and currently foreseen up to 2035, is expected to yield a total integrated luminosity of about 3000 fb$^{-1}$. 

\begin{figure}[hbtp]
\centering
\includegraphics[width=8cm, height=6cm]{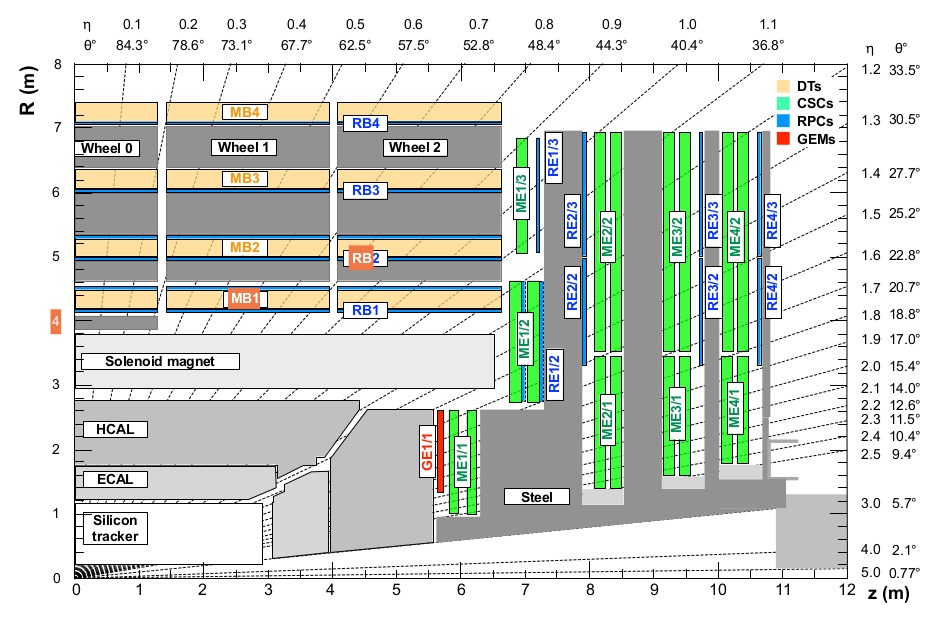} 
\caption{A quadrant of the R-z cross-section of the CMS detector, highlighting (in red) the location of the GE1$\slash$1 station in the pseudo-rapidity region $1.6<\left|\eta\right|<2.2$.} \label{fig:gem_system}
\end{figure}
During the HL-LHC period, the Compact Muon Solenoid (CMS) experiment~\cite{CMSDetector}  will continue its present physics program and therefore needs to maintain its  sensitivity for electroweak scale physics and for TeV scale searches.  As such, the CMS Collaboration is planning several detector upgrades in order
to maintain or improve its high level of  performance, in particular also of its muon system~\cite{CMSMuonperf}. Presently, as can be seen in
Figure~\ref{fig:gem_system}, only Cathode  Strip Chambers (CSC) are installed in the forward region $1.6<\left|\eta\right|<2.4$ of the CMS muon system. To increase redundancy and enhance the muon trigger and reconstruction capabilities in that particular region, several additional muon stations will be added~\cite{one_01}. A first step will be the installation during LS2 of an additional set of muon chambers denoted as  GE1$\slash$1 in the first muon endcap disks.  
The GE1$\slash$1 chambers\footnote{In "GE1$\slash$1", the "G" stands for GEM and the "E" for  Endcap; the first "1" corresponds to the first muon station and the second "1" to the first, innermost ring of the station} will contain Gas Electron Multiplier (GEM)  technology \cite{Sauli} which given its known excellent rate capability and radiation hardness is well-suited for this forward detector region. 

\section{GE1$\slash$1 motivation}
During the HL-LHC operation, the increase of the beam energy and the  collision rate compared to the current  situation will definitely affect the radiation environment of the detector.  First of all, the increase of the background rate in the forward region of the CMS muon endcaps will provoke a rise of the CMS
level-1 muon trigger rate and a degradation of the muon selection due to trigger bandwidth limitations. Therefore, the challenge associated with the forward region of the muon system for the HL-LHC is to maintain an efficient and reliable trigger in the $\left|\eta\right|>1.6$ region. High muon trigger rates in the forward region of the detector are driven by the fast drop in the magnetic field towards high $\left|\eta\right|$, which results in a decreased bending of muon trajectories as they traverse the CMS muon system. As the measurement of muon p$_{T}$ by the CMS level-1 muon trigger is effectively based on the observed bending of muon trajectories, any degradation in the muon p$_{T}$ resolution will lead to higher trigger rates due to the increased probability of low-p$_{T}$ particles being reconstructed as high-p$_{T}$ muons. 

In addition, the high radiation background may accelerate the aging of the current muon system and could cause performance losses and dead regions. The expected background in the CMS endcaps for an instantaneous luminosity of 5 $\times$ 10$^{34}$ cm$^{-2}$s$^{-1}$ has been computed using the  CMS adaptation of the FLUKA simulation package and was found to be dominated by neutrons and secondary particles arising from neutron interactions with matter. Neutrons with energies ranging from thermal values to a few GeV  originate from interactions of hadrons produced in primary pp collisions with the beam pipe material and structures in the very forward region. Neutron interactions within the detector material lead to secondary particles, mostly high-energy photons and electrons, that can produce  detectable amounts of ionization in gaseous detectors. The sensitivity of muon detectors to background particles with different energies has been computed using the GEANT4 framework~\cite{one_01}. The background spectrum has been convoluted with the sensitivity to provide the expected hit rate in the detectors. The maximum hit rate in the first muon station is expected to be around 5~kHz$\slash$cm$^{2}$ at the HL-LHC, with an estimated integrated charge of about 100~mC/cm$^2$ for 20 years of HL-LHC operation~\cite{one_01}. 

This implies that the upgraded forward muon system must be sufficiently resistant to radiation and have a high rate capability. Furthermore, it  must offer adequate pattern recognition capability to allow for  efficient reconstruction of muon tracks while minimizing the number of mis-identified tracks. Part of the solution chosen by the CMS Collaboration is the introduction of additional, GEM technology based gaseous detectors in the forward endcaps to complement the current CSC system.  The new GE1$\slash$1 station in the first endcap disks  together with the current muon system will enable the CMS trigger to better discriminate high-$p_{T}$ muons from low-$p_{T}$ muons~\cite{one_01}. In particular,  the GE1$\slash$1 station will extend the total path length of the muon system in that $\eta$-region and will provide additional  hits that will help to refine the stub reconstruction and improve the momentum resolution. With the new station installed, muon direction will be measured using hit positions in the adjacent GEM GE1$\slash$1 and CSC ME1$\slash$1 chambers in the same $\eta$-region. The good position resolution of both detectors and the increased  lever arm formed by the two detectors will allow for an improved measurement of muon direction and bending angle in the CMS magnetic field. The rate of misidentified muons as well as the L1 trigger rate will also be reduced. The reduction of trigger rate will allow maintaining a low momentum threshold, therefore increasing the acceptance and efficiency of reconstruction of soft muons.

\section{The GE1$\slash$1 station}

GEM detectors exploit the electron amplification that occurs within a gas medium inside narrow holes that perforate a thin polyimide foil in a triangular pattern. The polymide foil is clad on both sides with a thin conductive copper layer. With a voltage up to about 400 volts applied across the two copper-clad surfaces of a foil a strong electric field (60-100 kV$\slash$cm) is produced inside the GEM holes. The primary electrons produced due to ionization of the gas 
by a  charged particle passing through the chamber  drift towards the holes and once they start experiencing the very intense electric field inside the holes, they acquire enough kinetic energy to produce secondary ionization in the gas. This process eventually leads to the formation of an electron avalanche. An arrangement of three cascaded GEM  foils, commonly known as a "triple-GEM detector", allows a high charge amplification factor up to several 10$^{5}$ for modest applied high voltage, which limits the probability of electrical breakdown. The amplified charge induces a measurable signal 
on a readout electrode that can be segmented to provide positional information. 

The GE1$\slash$1 detectors are trapezoidally shaped and consist of a gas volume containing a  stack of three large-area GEM foils, i.e. a triple-GEM detector, embedded between a drift electrode and a readout board, with an induction/transfer-1/transfer-2/induction gap configuration of 3$\slash$1$\slash$2$\slash$1~mm. The first transfer gap has been set in order to minimize the charge released after the first GEM. The induction gap has been set to achieve induction field up to
5~kV$\slash$cm without too much increase in the potential across the entire structure.  The baseline gas mixture for operating the CMS triple-GEMs is Ar/CO$_{2}$ in 70:30 proportion.

\begin{figure}[hbtp]
\centering
\includegraphics[width=8cm, height=6cm]{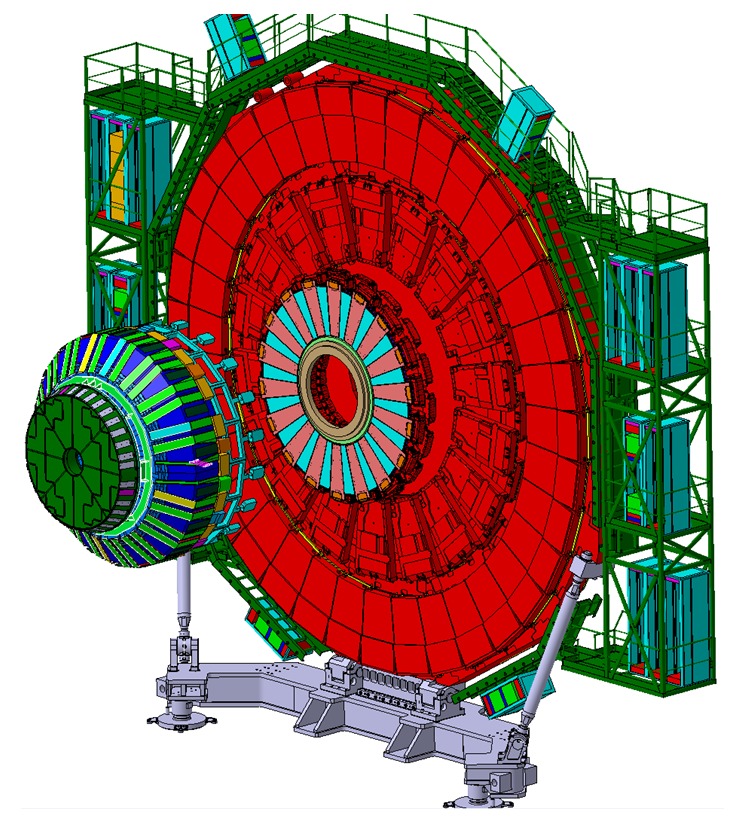}
\caption{\label{fig:ge11_station}The first CMS muon endcap station with the GE1$\slash$1 super-chambers in the inner ring.}
\end{figure}
  
In the GE1/1 stations, a pair of triple-GEM detectors is combined to form a "super-chamber" that provides two measurement planes to maximize detection
efficiency. Each super-chamber covers a 10.15$^{\circ}$ sector so that 36 super-chambers are required to form a ring that gives full azimuthal coverage as can be seen in Figure~\ref{fig:ge11_station}. To instrument both endcap disks, i.e. two GE1/1 stations, a total of 72 super-chambers or 144 basic chamber units are required.  The super-chambers will alternate in phi between long and short versions as dictated by the mechanical envelope of the existing endcap disk, i.e. each endcap disk will hold 18 long and 18 short super-chambers. 

Over the past years, the performance of several generations of GE1$\slash$1 chamber prototypes was studied in a series of beam tests at CERN and Fermilab~\cite{one, two_01}. With 98.0$\%$ chamber efficiency, a GE1$\slash$1 super-chamber will have an  efficiency above 99.9$\%$ when the logical OR of the signals from the two basic units is taken.

\section{Chamber design}

\begin{figure}[hbtp]
\centering
\includegraphics[width=8cm, height=6cm]{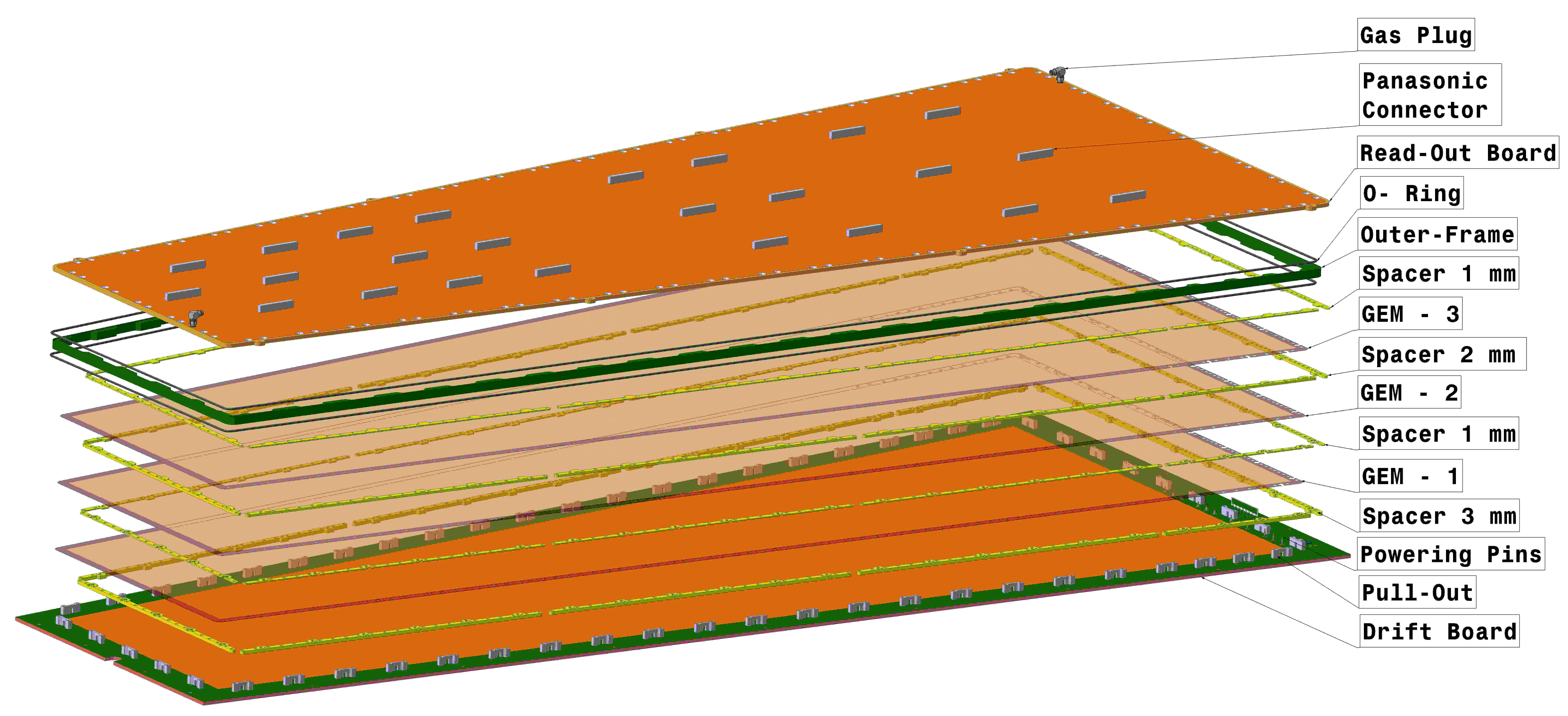}
\caption{GE1$\slash$1 layout and its main components starting from bottom: drift board mounted all around with stainless steel pull-outs used for stretching of GEM foils, 3 mm frame (Spacer), first foil, 1 mm frame, second foil, 2 mm frame, third foil, 1 mm frame, first O-ring, external frame, second O-ring and the readout board.} \label{fig:gem_detector}
\end{figure}

The structure of a basic GE1/1 unit is shown in Figure~\ref{fig:gem_detector}. A chamber consists of a drift board, three identical GEM foils stacked together within frames, a readout board and an external gas frame. The drift and readout boards and the external frame define the gas  volume with the gas tightness ensured by an O-ring placed inside a groove in the  external frame. This construction allows the detectors to be thin, which is important given the limited space available within the existing CMS detector. Because of mechanical constraints due to support structures in the GE1/1  station, two versions  of detectors were designed in order to maximize detection coverage. Long chambers have a radial length of 128.5~cm, while short chambers have a radial length of 113.5~cm. The main technical specifications of the GE1/1 detectors for both the Long and Short versions are listed in Table~\ref{table:1}.

\begin{table}[ht]
\centering
\scriptsize
\begin{tabular}{|p{2.7cm}||p{2.1cm}|p{2cm}|}
\hline
\multicolumn{3}{|c|}{GE1$\slash1$ detector} \\
\hline	
\hline			
Specification	&Short	&Long \\		
\hline	
\hline
Shape	&Trapezoidal	&Trapezoidal\\
Chamber Length	&113.5 cm	&128.5 cm\\
Chamber Width	&(28.5-48.4) cm	&(28.6-51.2) cm\\
Chamber thickness	&1.42 cm	&1.42 cm\\
Active readout area	&3787 cm $^{2}$ (app.)	&4550 cm$^{2}$ (app.)\\
Active chamber volume	&2.6 liters	&3 liters\\
Geometric acceptance in $\eta$	&1.61-2.18	&1.55-2.18\\
\hline
\end{tabular}
\caption{Technical specifications of the GE1$\slash$1 Short and Long chambers.} \label{table:1}
\end{table}

The design of the drift board, the external and internal frames, the GEM foils, the readout board and the gas distribution system are described in the following subsections.  

\subsection {The drift board}  \label{drift_Board}
\begin{figure}[hbtp]
\centering
\includegraphics[width=5cm, height=3.3cm]{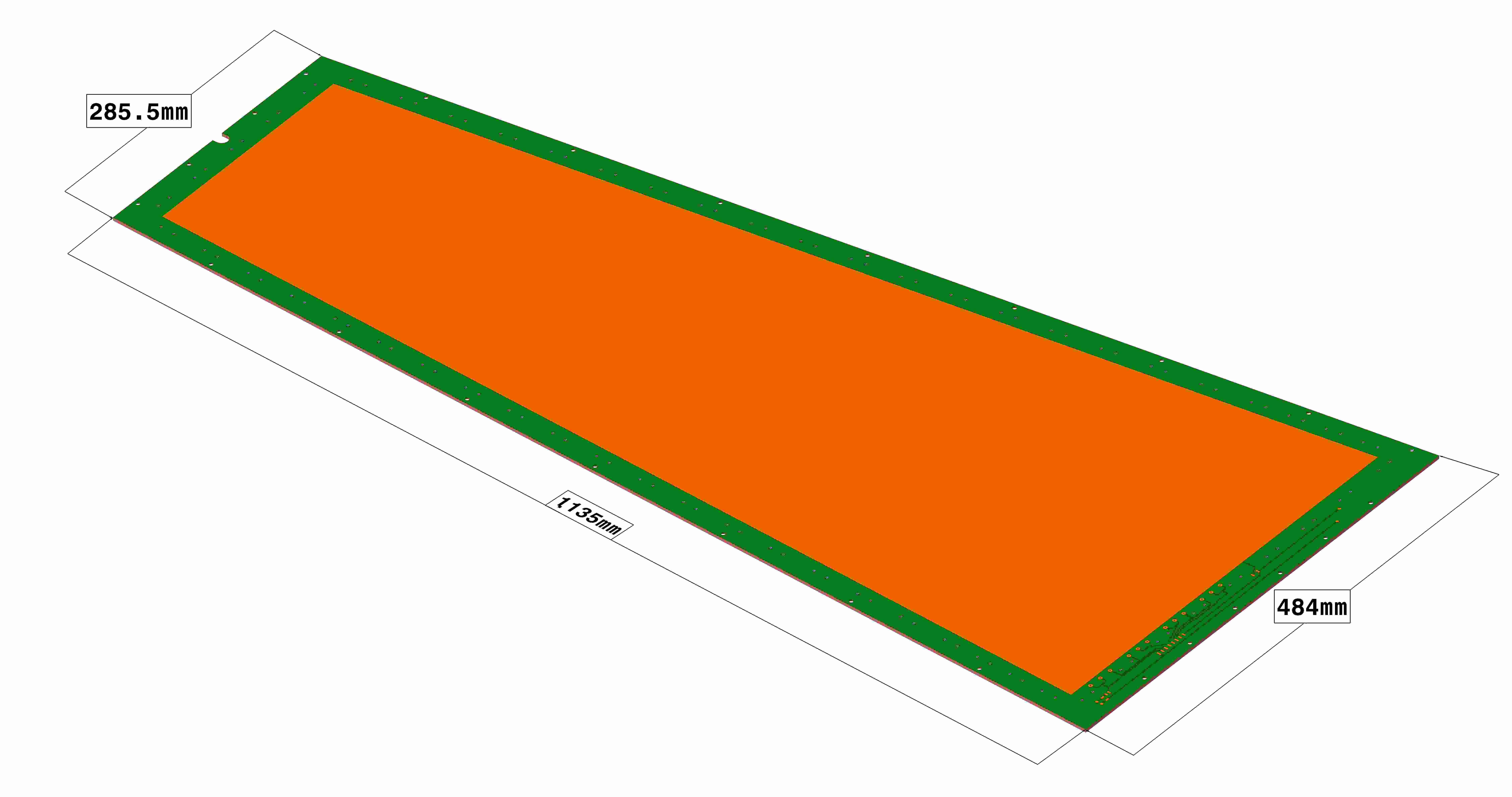}  
\includegraphics[width=3.5cm, height=2.5cm]{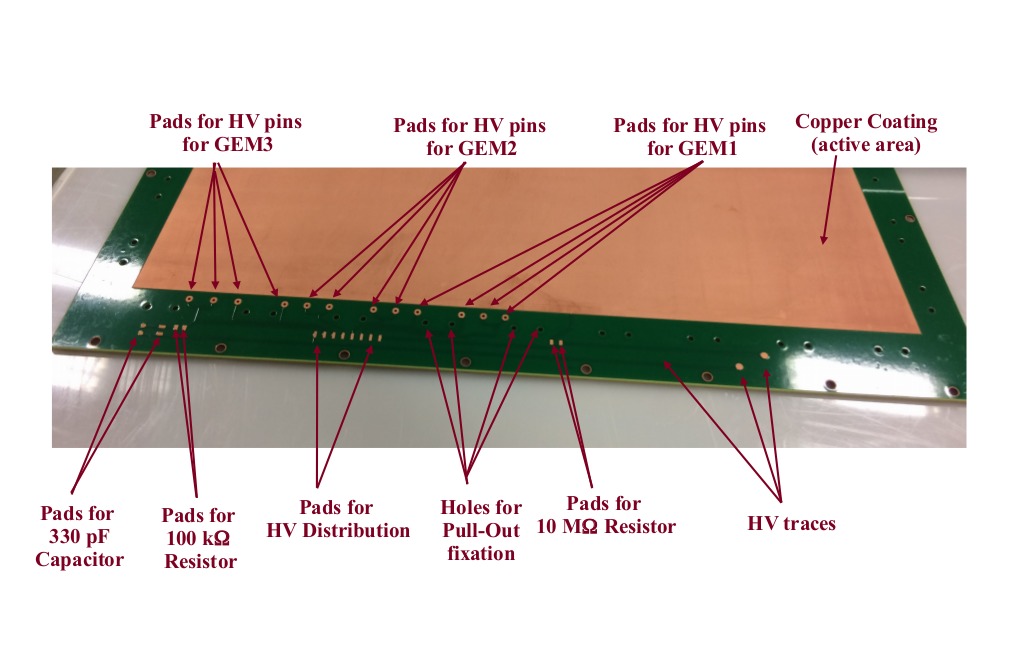}
\caption{Design of GE1$\slash$1 drift board (left) and close-up view of the wide end of  the actual GE1/1 drift board (right) showing the on-board HV circuit traces and pads
for the  spring-loaded pins that make the electrical connections to the GEM foils; the pads for a 10~M$\Omega$ SMD resistor, 100~k$\Omega$ and 330~pF capacitor; the holes to fix pull-outs against the board.}
\label{fig:drift} 
\end{figure}

\begin{figure}[hbtp]
\centering
\includegraphics[width=8cm, height= 4cm]{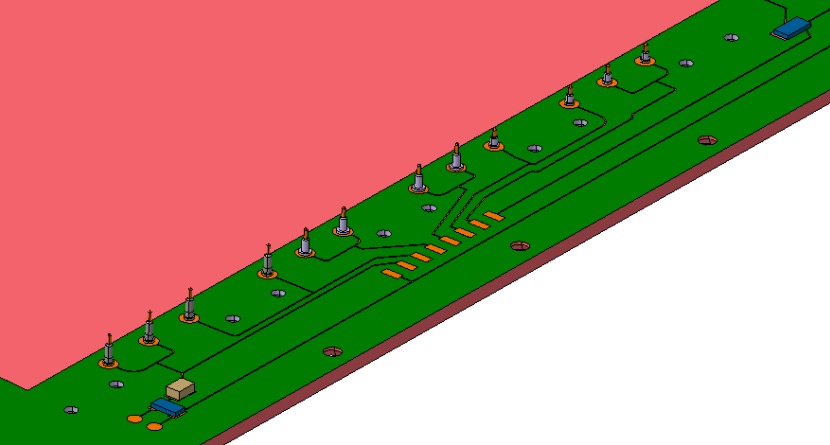}  
\caption{Magnified view of the design of the drift board, with the twelve HV pins and corresponding  soldering pads for the resistive divider network.} 
\label{fig:Power_Pins_design}
\end{figure}

The GE1/1 drift board is a trapezoidal-shaped printed circuit board holding the drift electrode. The actual board and the magnified view of its wider side is shown in Figure~\ref{fig:drift}. The board has an active area coated with a copper layer, i.e. the drift electrode, that is contained in the active gas volume. GEM  foils are electrically connected to the HV power supply using spring-loaded pins. As shown in Figure~\ref{fig:Power_Pins_design},  these HV pins are mounted on the drift board, at positions corresponding to the HV pads located on the GEM foils. Four pins are foreseen for each foil, i.e. a bottom and top side contact plus two spare pins. Pins for a  given GEM foil, i.e. at a given height in the stack, have an identical height  height. 
The drift board has number of holes around the periphery to accommodate mechanical stainless steel pieces called pull-outs, that are used to stretch the foils as will be explained in Section~\ref{stretching_Mechanism}. 

Other on-board elements are outside the gas volume. There are pads for a 10~M$\Omega$ protection resistor that is used to limit the current from the HV  power supply. There are pads for a 100~k$\Omega$ resistor and a 330~pF capacitor used to  decouple the signal from the HV when during the post-assembly chamber quality
control  procedure a detector signal is extracted from the bottom of the third GEM foil. Finally, there are HV traces as shown in Figure~\ref{fig:drift}.

\subsection {The external frame}

\begin{figure}[hbtp]
\centering
\includegraphics[width=3cm, height=4.2cm]{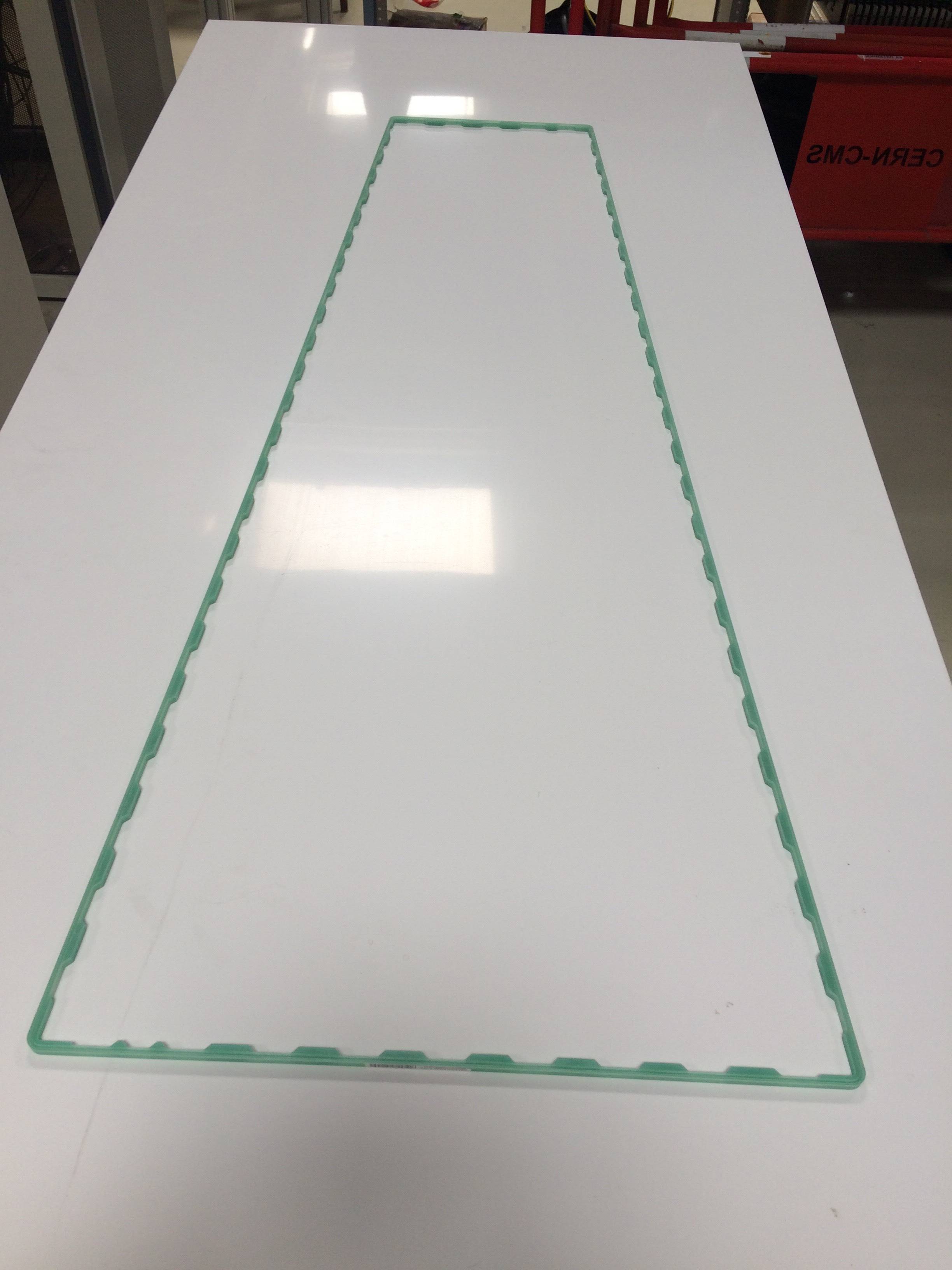}  
\includegraphics[width=4.5cm, height=4cm]{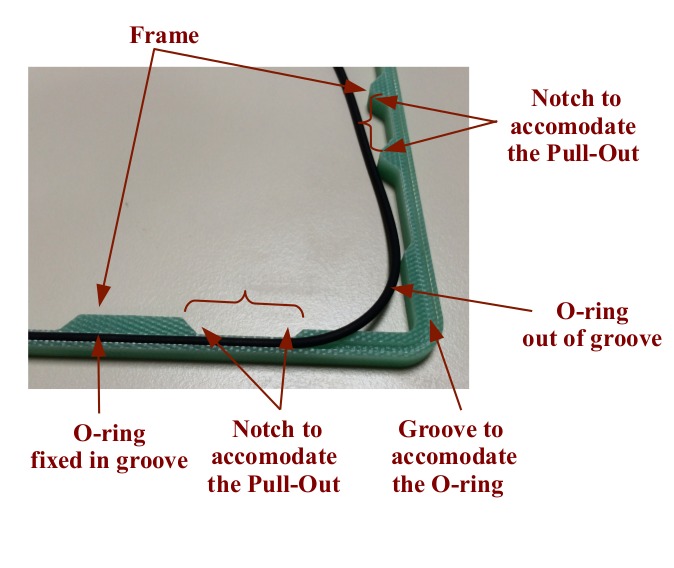}
\caption{(left) The GE1/1 epoxy-glass external frame. (right) Close-up view   of the section of the external frame showing the groove in the frame, O-ring
in and out of the groove and notches in the inner side of the frame to accommodate the pull-outs.} 
\label{fig:ex_frame} 
\end{figure}

The GE1/1 external frame is shown in Figure~\ref{fig:ex_frame}. It is made up of halogen-free glass-epoxy material, machined from a single piece. The frame is used to close the active gas volume between the drift and readout boards. It has a trapezoidal shape and has numerous wide notches to accommodate the stainless steel pull-outs. It is coated with Nuvovern polyurethane varnish before assembly to seal in particulates. Narrow grooves
are machined on both sides in order to accommodate a Viton O-ring. 

\subsection {The internal frame} \label{Section_Internal_frames_design}
\begin{figure}[hbtp]
\centering
\includegraphics[width=7cm, height=6cm]{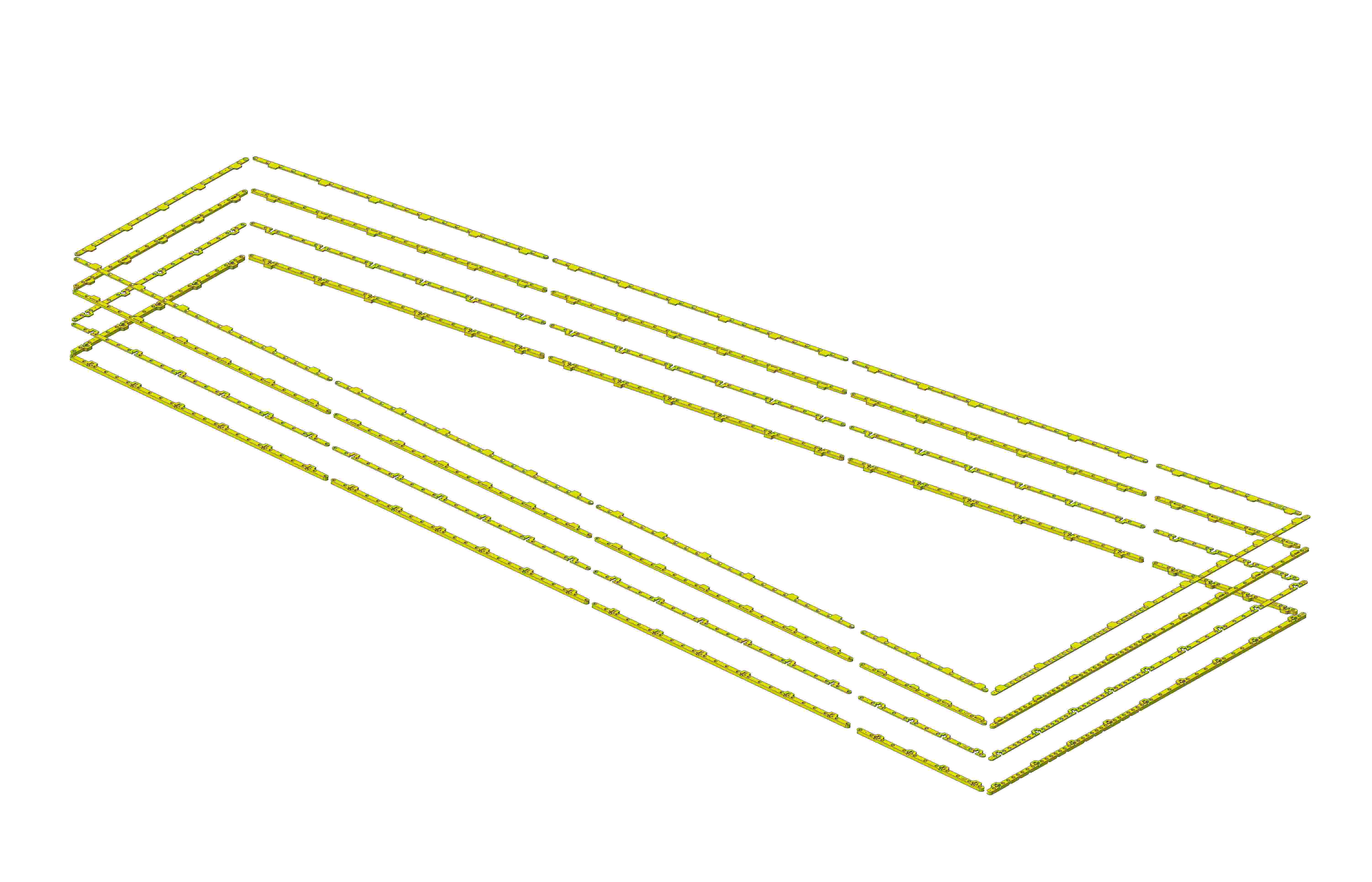} 
\caption{Design of the GE1/1 internal frames.} 
\label{fig:Internal_frames_design} 
\end{figure} 

The design of the internal frames is shown in Figure~\ref{fig:Internal_frames_design}. There are four layers of internal frames made from halogen-free glass epoxy with thickness of 3~mm, 1~mm, 2~mm, and 1~mm. Each frame is composed of 10 individual pieces per layer. The pieces are coated with Nuvovern polyurethane varnish before assembly which ensures that no glass epoxy particulates get detached from the frames during the chamber assembly as any dust falling onto the GEM foils could produce electrical  shorts in the holes. Small threaded M2 brass inserts are fixed within the 3~mm frame to avoid loosening of macroscopic and microscopic glass epoxy particulates from the frames when screws pass through the frame. The frame layers are stacked and define the 3$\slash$1$\slash$2$\slash$1~mm spacings between the drift board, the triple-GEM structure and the readout board.
 
\begin{figure}[hbtp]
\centering
\includegraphics[width=3.3cm, height=3cm]{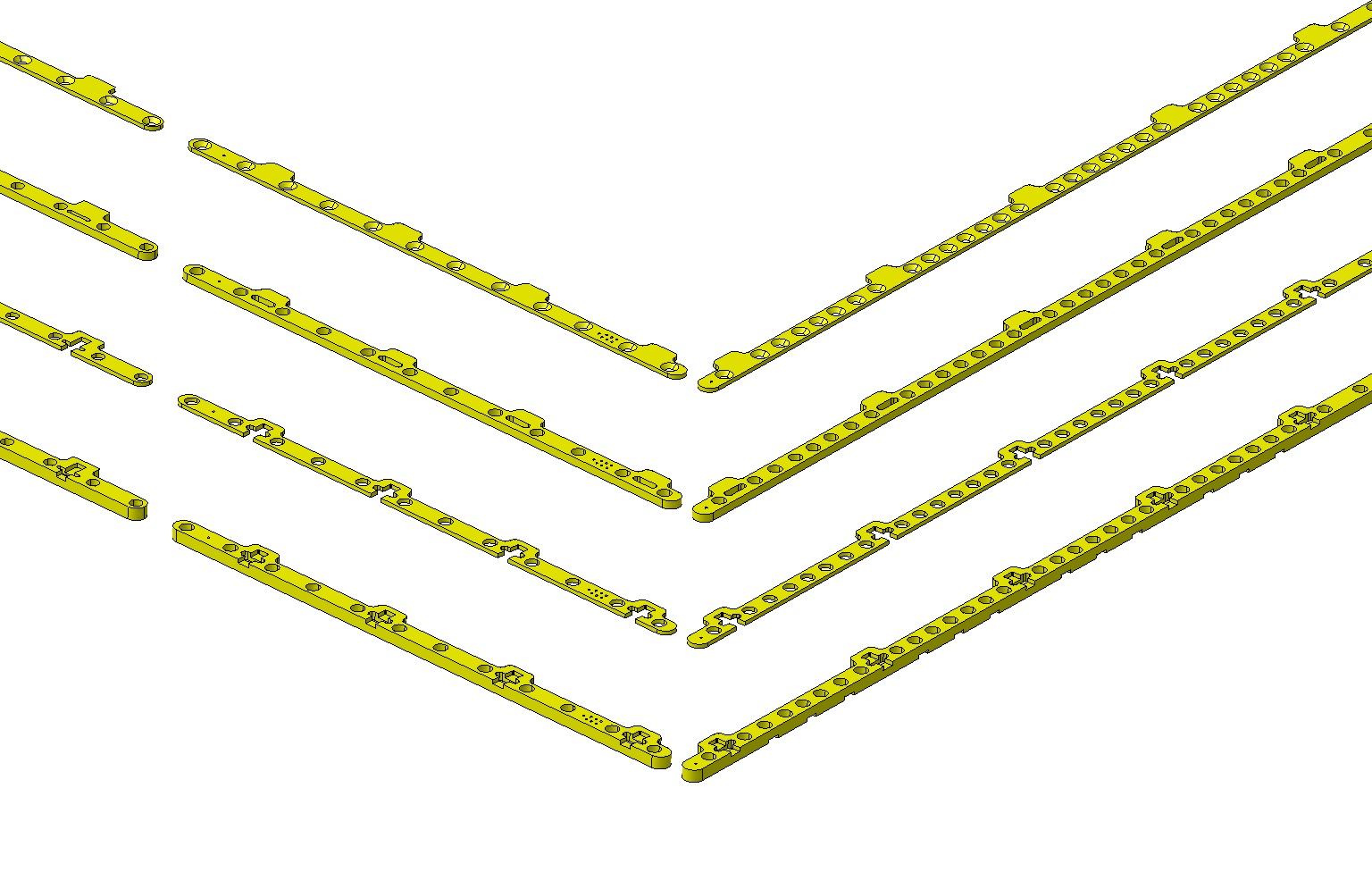}
\includegraphics[width=3.3cm, height=3.2cm]{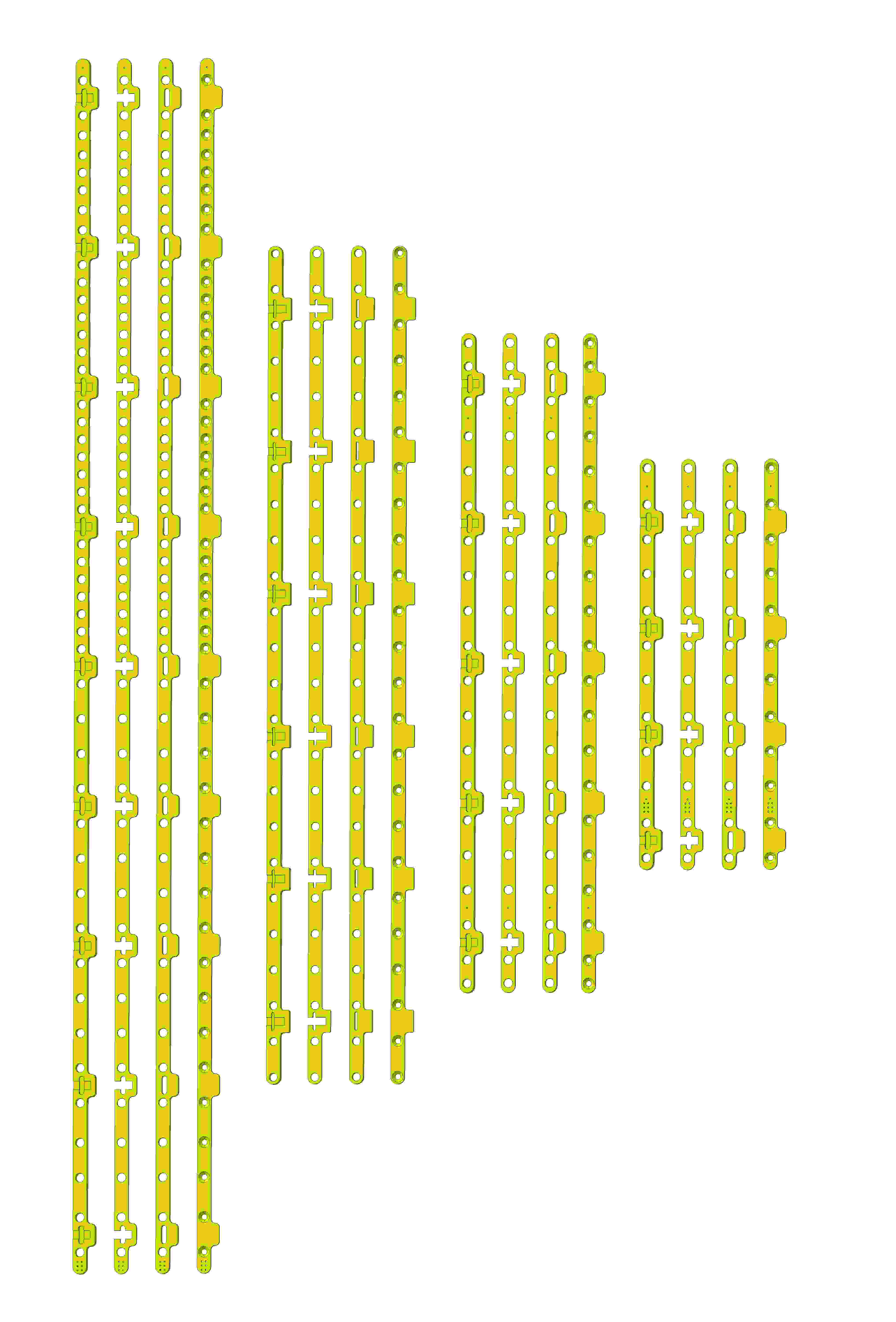}  
\caption{Magnified view of a section of the Figure~\ref{fig:Internal_frames_design} (left) and shapes and mechanical structure of the different pieces of the frames (right). Ten pieces are joined together to form each frame. The stacked frames surround the GEM foils along their periphery.} 
\label{fig:Internal_frames}
\end{figure} 
  
\subsection {The GEM foils} 
\begin{figure}[hbtp]
\centering
\includegraphics[width=4cm, height=5.1cm]{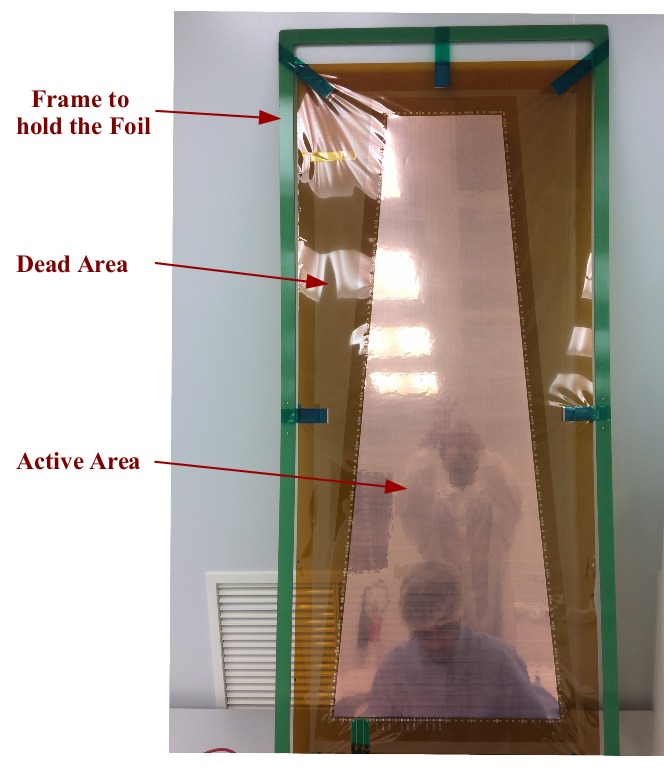}  
\includegraphics[width=4cm, height=5cm]{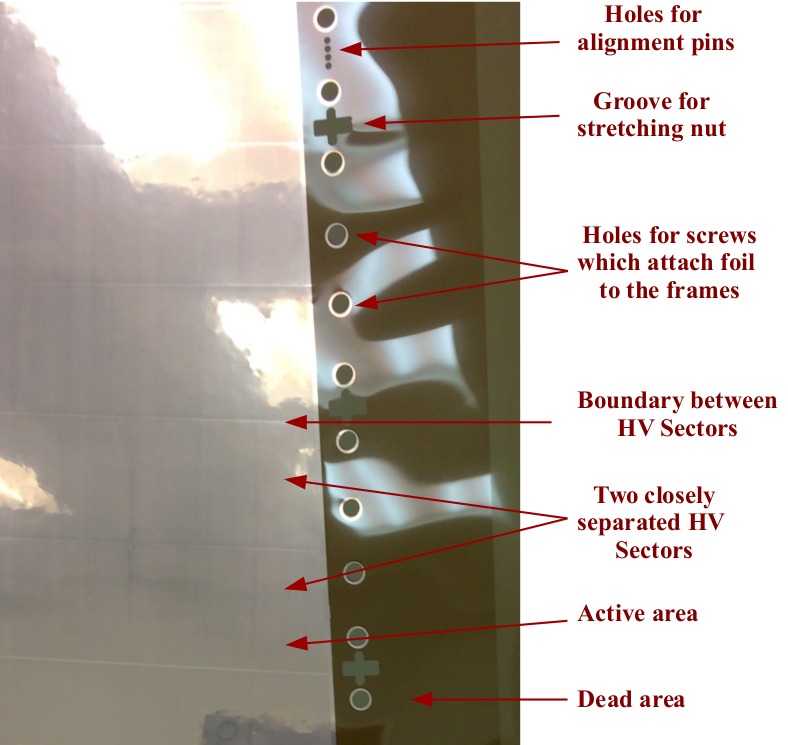} 
\caption{Design of the GE1/1 GEM foils (left) and close-up view of a section of a foil (right). Small holes for alignment pins are used during assembly. Large holes allow the passage of screws used to attach a foil to the internal  frame. "Plus"-shaped grooves are there to accommodate stretching nuts. The
boundaries between  the different HV sectors are visible.} 
\label{fig:GEM_FOil}
\end{figure}

The GE1$\slash$1 detector uses three identical trapezoidal-shaped GEM foils as shown in Figure~\ref{fig:GEM_FOil}. The foils are produced at the CERN PCB workshop using a single-mask production technique \cite{two_01a}. The GEM foil surfaces oriented towards the readout board are a single continuous conductor whereas the GEM foil surfaces oriented towards the drift board are segmented into sectors. The sectors run across the trapezoid in the same direction as the parallel ends. The sector width (in the radial direction) is largest at the short end of the trapezoid and smallest at the wide end so that the area of each sector is approximately the same, about 100~cm$^{2}$. This segmentation limits the charge and energy in case of a discharge. In an extreme case,  if a discharge large enough to generate a short were to occur in a particular HV segment, it would render unusable only that particular HV segment and not the entire  foil. Therefore, each segment has a separate connection to the HV supply via a trace around the edge of the GEM foil to a common connection point at the wide end. Each
trace is connected through a 10~M$\Omega$ surface-mounted protection resistor that limits the current from the HV supply, decouples the capacitance from
other HV sectors, and quenches a discharge. The voltage for the other side of the foil is connected at two points located at the wide end of the foil, where the two points provide redundancy. 

\subsection {The readout board}
\begin{figure}[hbtp]
\centering
\includegraphics[width=7cm, height=4.5cm]{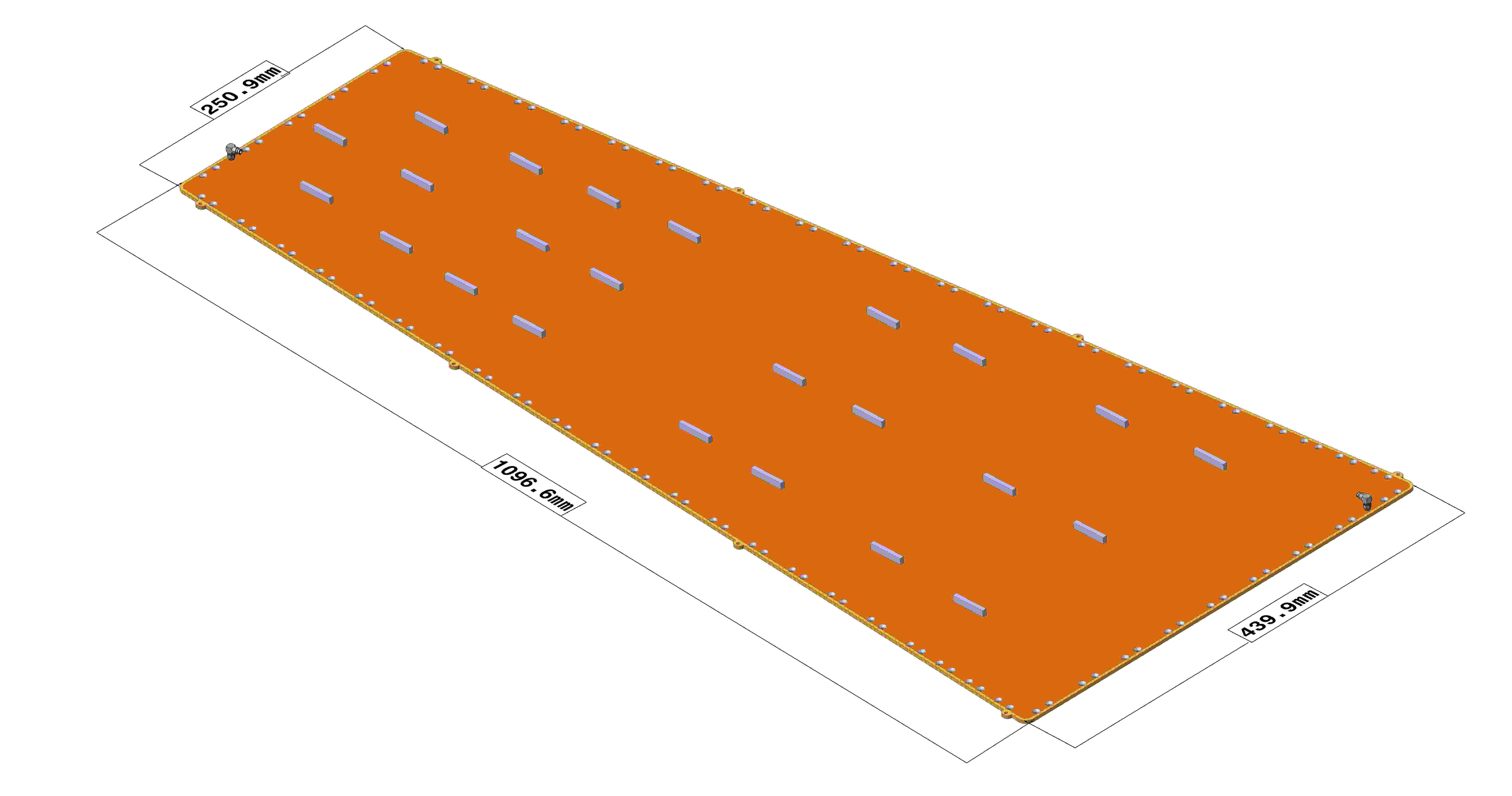}  
\caption{Design of the readout board with the gas plugs fixed at opposite
edges.} 
\label{fig:readout_design}
\end{figure}

The readout board is a trapezoidal-shaped printed circuit board with mechanical design as shown in Figure~\ref{fig:readout_design}. The inner side of the board features 3072 trapezoidal readout strips oriented radially along the longer sides of the chamber. The active area covered by the strips subtends an azimuthal angle of 10.15$^{\circ}$ which allows for an overlap of 0.15$^{\circ}$ (equivalent to 5.67 strips) between the active areas of adjacent chambers. All the readout strips are connected through metalized vias to the outer side of the board where traces are routed from the vias to readout pads in partitions of 8 $\times$ 3 partitions in ($\eta$, $\phi$). Each $\eta$-partition has 384 strips comprised of three 128-strip sectors  in $\phi$. The strip pitch varies between 0.6~mm at the shorter end of the chamber and 1.2~mm at the wider end. The readout board has two holes at diagonally opposed corners, in which gas plugs are mounted that serve as inlet and outlet explained in Section~\ref{Gas_Distribution}.  

\begin{figure}[hbtp]
\centering
\includegraphics[width=2.3cm, height=4.5cm]{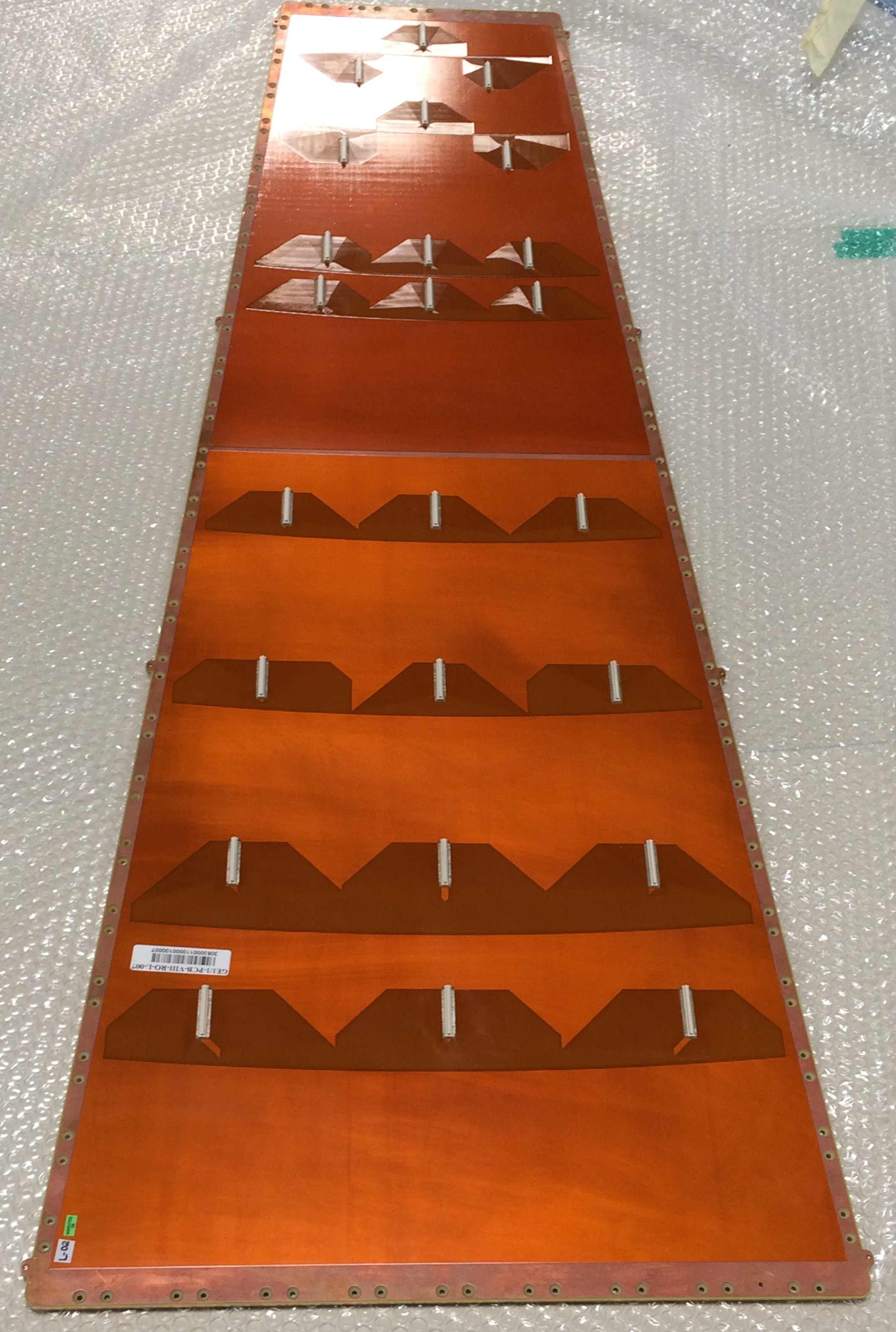} 
\includegraphics[width=2.3cm, height=4.5cm]{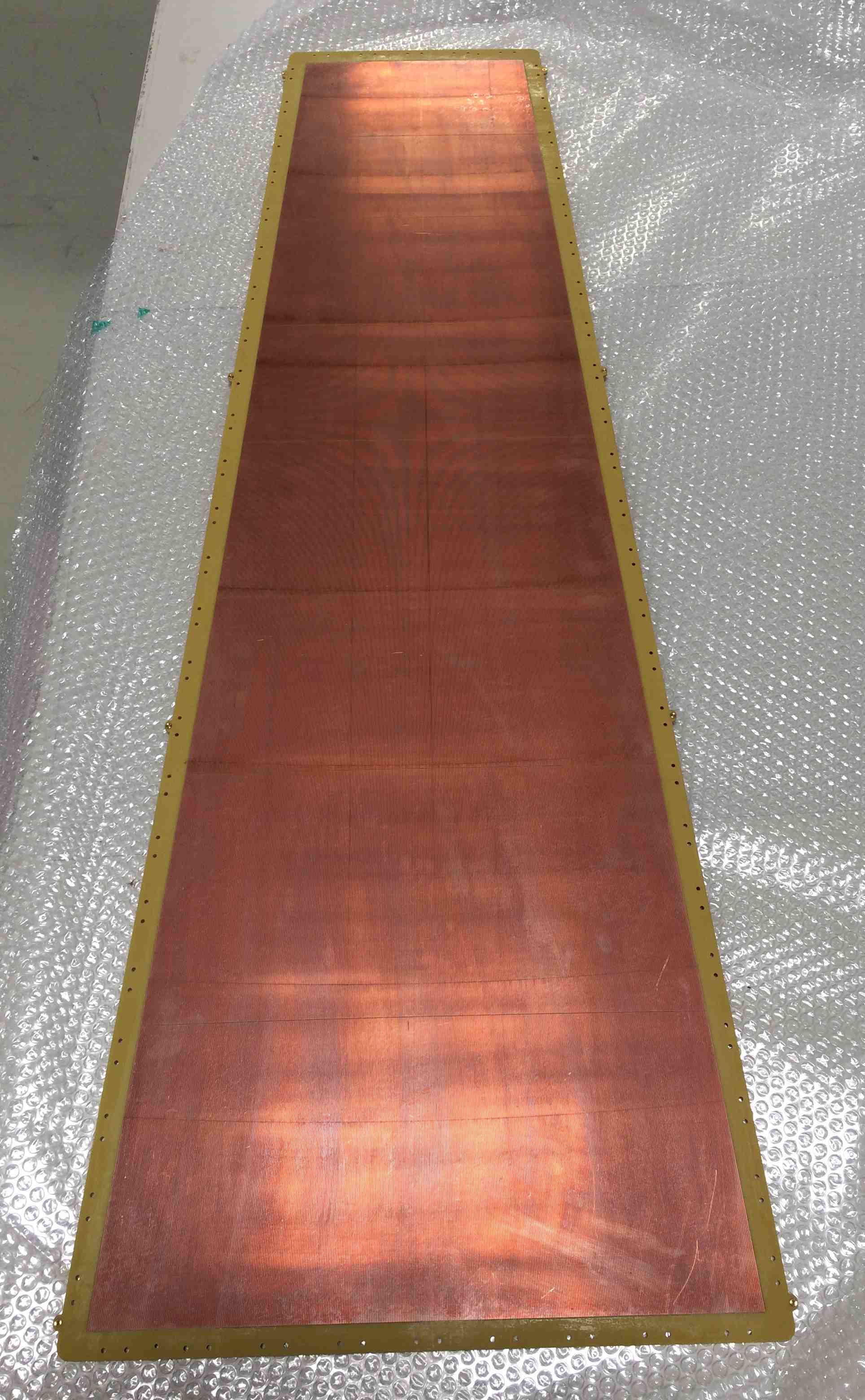}
\includegraphics[width=2.3cm, height=4.5cm]{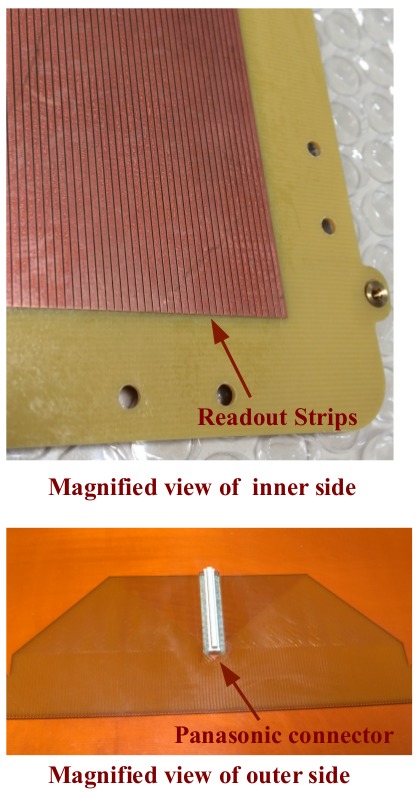}  
\caption{(Left) Outer side of GE1$\slash$1 readout board showing 24 ($\eta$, $\phi$) readout sectors, each with a male Panasonic connector for signal  readout. There are holes at opposite corners used to mount the two gas plugs; one serves as gas inlet and another as gas outlet. Holes on the periphery of the board allow the passage of screws that fix the board against the chamber structure. (Middle) Inner side of the drift board showing the readout strips. (Right) Close-up view showing readout strips on the inner side and a Panasonic connector on the outer side of the board.} 
\label{fig:readout}
\end{figure}

\subsection{On-chamber HV distribution to the GEM foils and drift electrode}
\label{Divider}

\begin{figure}[hbtp]
\centering
\includegraphics[width=7cm, height= 4 cm]{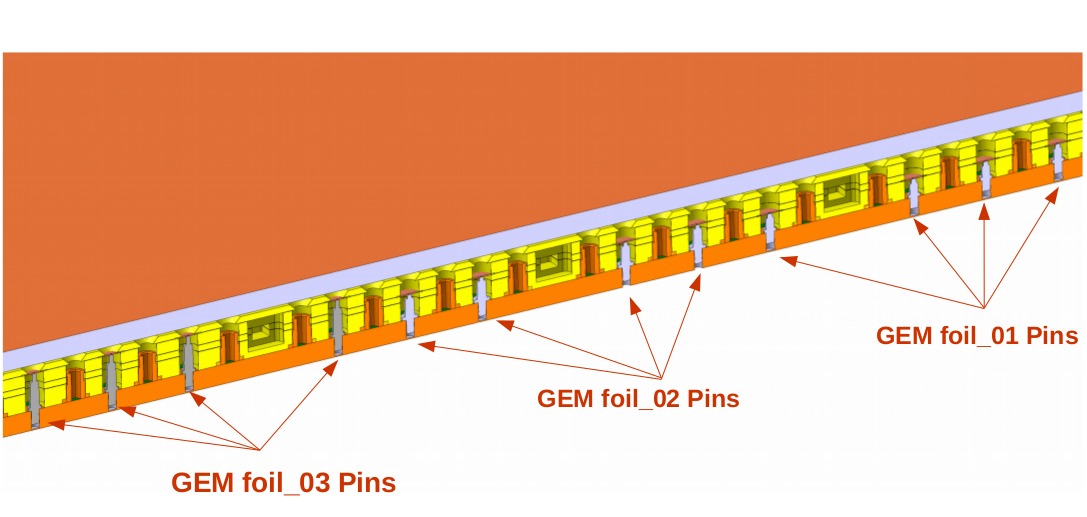}  
\caption{Sliced view of the chamber design showing the mechanism to power the three  GEM foils with twelve HV pins.} 
\label{fig:Power_design}
\end{figure}

\begin{figure}[hbtp]
\centering
\includegraphics[width=7cm, height= 4cm]{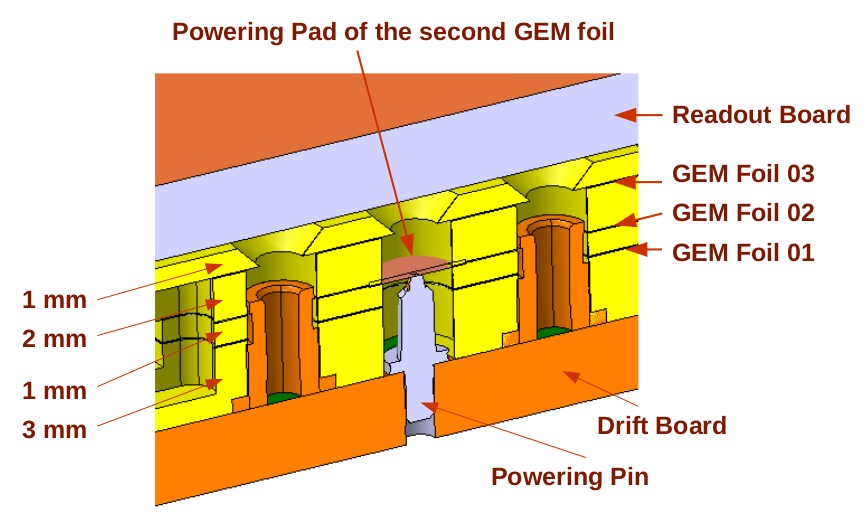} 
\caption{Magnified sliced view of Figure~\ref{fig:Power_design} showing various components such as the drift board, 3 GEM foils in a stack, 3~mm, 1~mm, 2~mm, 1~mm frames, second GEM foil powering pad, readout board and the single HV pin pressed against  powering pad of the second GEM foil.}  
\label{fig:Single_Power_design}
\end{figure}

\begin{figure}[hbtp]
\centering
\includegraphics[width=5.3cm, height= 2.7cm]{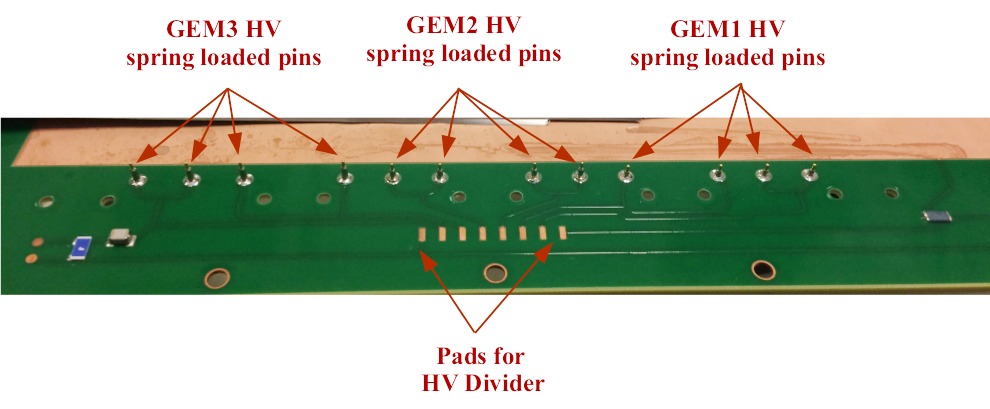} 
\includegraphics[width=2.3cm, height= 3.7cm]{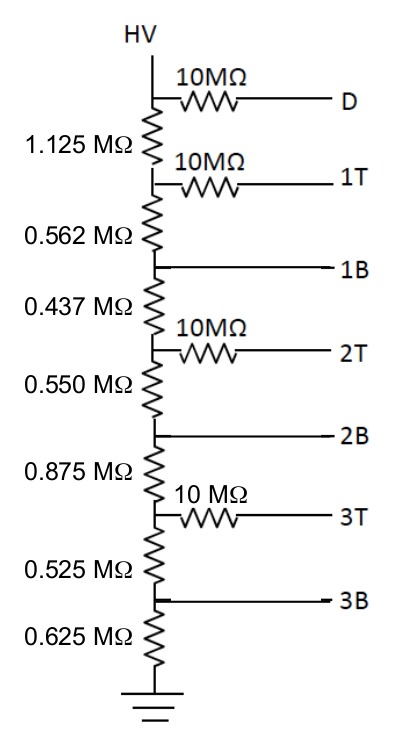} 
\caption{(left) Twelve spring-loaded pins soldered onto the drift board to make electrical HV connections to corresponding contact pads on the GEM foils; the
three sets of pins have different heights so that they can properly reach the three GEM foils. (right) The resistive divider network used to power up the
GE1$\slash$1 detector with a gap configuration of  3$\slash$1$\slash$2$\slash$1~mm using single channel power supply; the notation D, 1T, 1B, 2T, 2B, 3T  and 3B corresponds to the drift board, the top and bottom electrodes of the first, second and  third GEM foil respectively.} 
\label{fig:Divider}
\end{figure}

As described in Section~\ref{drift_Board}, the GEM foils are powered through high voltage pins soldered onto the drift board within the gas volume, the design of which is shown in Figure~\ref{fig:Power_design}, while an actual photograph is shown in Figure~\ref{fig:Divider}. These pins get pushed against their corresponding connection pads on the GEM foils which are at different heights as shown in Figures~\ref{fig:Power_design} and \ref{fig:Single_Power_design}. The HV
pins are connected with pads outside the gas volume. These pads are designed in order to allow to power the GEM chambers either using multi-channel or single channel power supply. In the later case, the pads will be used to fix a special resistive divider network which allows a correct voltage distribution to the GEM foils. The circuit diagram for the 3$\slash$1$\slash$2$\slash$1~mm  chamber gap configuration is shown in Figure~\ref{fig:Divider}. 

\subsection{Gas distribution within the chamber} \label{Gas_Distribution}

\begin{figure}[hbtp]
\centering
\includegraphics[width=6.5cm, height= 4.5cm]{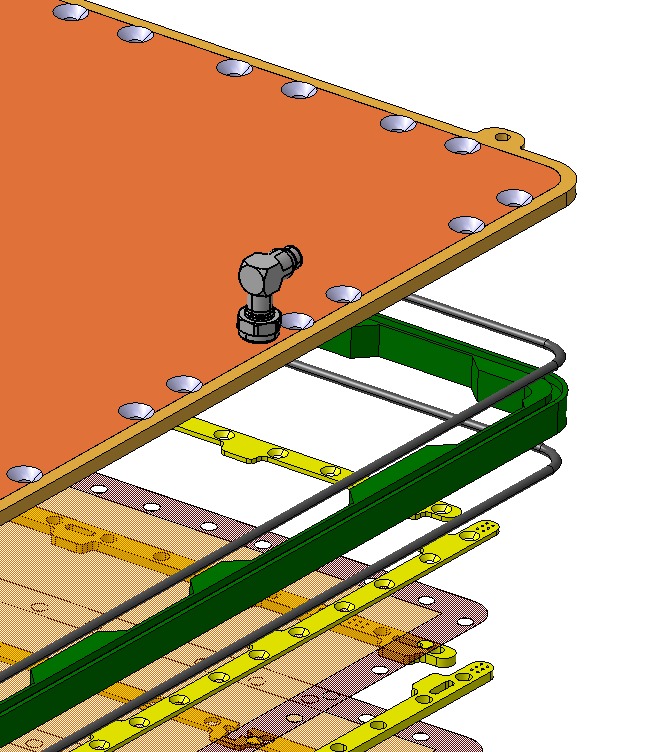} 
\caption{Exploded view of a GE1/1 chamber, including also the design of a gas plug fixed onto a GE1/1 readout board.} 
\label{fig:Gas_plug}
\end{figure}

Gas connections to each GE1/1 chamber are made with a single inlet and a single outlet located on  diagonally opposite corners of the readout board. The design of a gas plug fixed on the readout board is shown in Figure~\ref{fig:Gas_plug}. The gas mixture flows diagonally through the chamber and the presence of the GEM stack in the gas volume  (i.e., between the Readout, external frame, and Drift boards) directs the gas flow towards the other corners and makes the flow laminar. The gas distribution inside the GEM stack (i.e. between GEM foils) is made possible by diffusion through the gaps between the inner frames and through the GEM
holes. 

\section{Assembly technique}

Any kind of contamination can make a GE1$\slash$1 detector fail as  particles going inside the 10 micron diameters holes of the GEM foils can cause electrical shorts. Therefore, the assembly of such detectors takes place in a  clean room environment of ISO class 6. The use of tools with lubricated shafts,
soldering  equipment that requires heating of volatile fluxes, motors and vacuum pipes with out-gassing oils, hair and
fingernail cosmetics are sources which can possibly contaminate the detector and are forbidden in the assembly area. 

For the construction of the very first GE1/1 prototype, the GEM foils were initially  thermally stretched during 24 hours at 37$^{\circ}$ in a special oven~\cite{one_02}. Fiberglass spacer frames were then glued onto the foils in order to fix and keep them separated at the correct distance in the triple-GEM configuration. This procedure was however prone to possible glue contamination, very time-consuming and labour-intensive and therefore not well suited  for mass production. This glueing assembly procedure was eventually abandoned after the construction of the first  two generations of GE1$\slash$1 chambers. Alternative  methods were tested to stretch  GEM  foils including stretching due to infrared  heating lamps~\cite{one_03}. However, a major development in the GE1$\slash$1 chamber assembly procedure was introduced in 2011 with a new technique~\cite{one, two} in which the GEM foils are mechanically stretched and the chamber is constructed without the use of any glue. The technique was initially tested on small 30$\times$30~cm$^2$ prototypes~\cite{three} and
later used to construct actual GE1/1 chambers~\cite{four, five}. Since  no glue is needed during the assembly this technique reduces the chamber
assembly time  from several  days to a few hours only, and also allows for a chamber to be reopened in case of problems.  

The important steps performed during the assembly of a GE1$\slash$1 chamber are discussed in the forthcoming sections. The procedure starts with the preparation of the drift board, followed by the assembly of the GEM stack. Next, the GEM foils are mechanically stretched using the above mentioned new technique after which the chamber is closed by the readout board.   

\subsection {Drift board preparation} 
\label{drift_board}

The drift board is prepared outside the clean room in order to avoid the contamination of the clean room due to the soldering of different components which must be mounted on the board. Twelve pins are mounted on the drift board for making HV connections to the corresponding GEM foils. The three sets of pins have different heights to properly reach the three GEM foils in the stack.  Pull-outs are bolted onto the perimeter of the drift PCB with two $A2$ stainless steel M 3 $\times$ 6/ $\times$ 8 screws that are sealed with polyamide washers against the drift board to prevent gas leakage. As described already in Section~\ref{drift_Board}, a surface mounted (SMD) 10~M$\Omega$ Resistor, a 330~pF Capacitor and  a 100~k$\Omega$ resistor are also soldered to the drift board.

\subsection {GEM stack assembly} \label{stack}
\begin{figure}[hbtp]
\centering
\includegraphics[width=6.5cm, height=3.5cm]{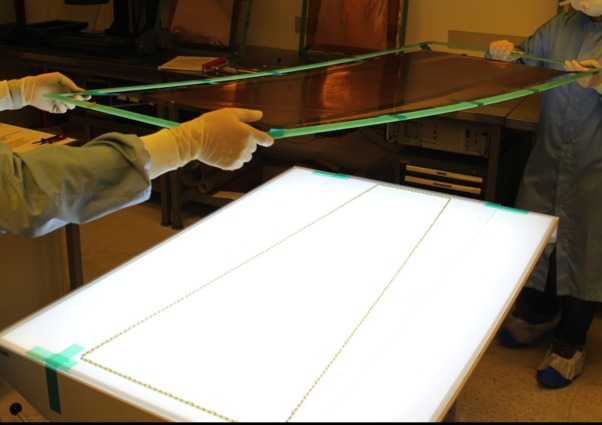} 
\caption{The 3~mm frame pieces placed on the Plexiglas plate forming the
trapezoidal shape of the chamber.} 
\label{fig:frame_foil}
\end{figure}

The most critical part in the chamber construction is the preparation of the GEM foil stack,  which is realized with the help of a Plexiglas base plate that includes  alignment pins at  desired positions to keep the entire structure aligned during the assembly procedure. These alignment pins are fixed into the Plexiglas sheet following the trapezoidal geometry demanded by  the mechanical layout of the detector. In the stack, the internal frame pieces fabricated in FR4 material are alternated with the GEM foils to define the different gaps. The 10 pieces of the 3~mm frames are inserted first over the alignment pins before placing the first GEM foil as shown in Figure~\ref{fig:frame_foil}.   

\begin{figure}[hbtp]
\centering
\includegraphics[width=3.8cm, height=2.9cm]{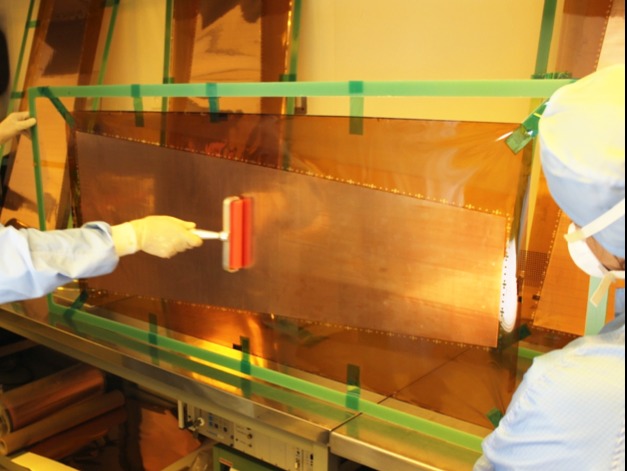}
\includegraphics[width=3.8cm, height=2.9cm]{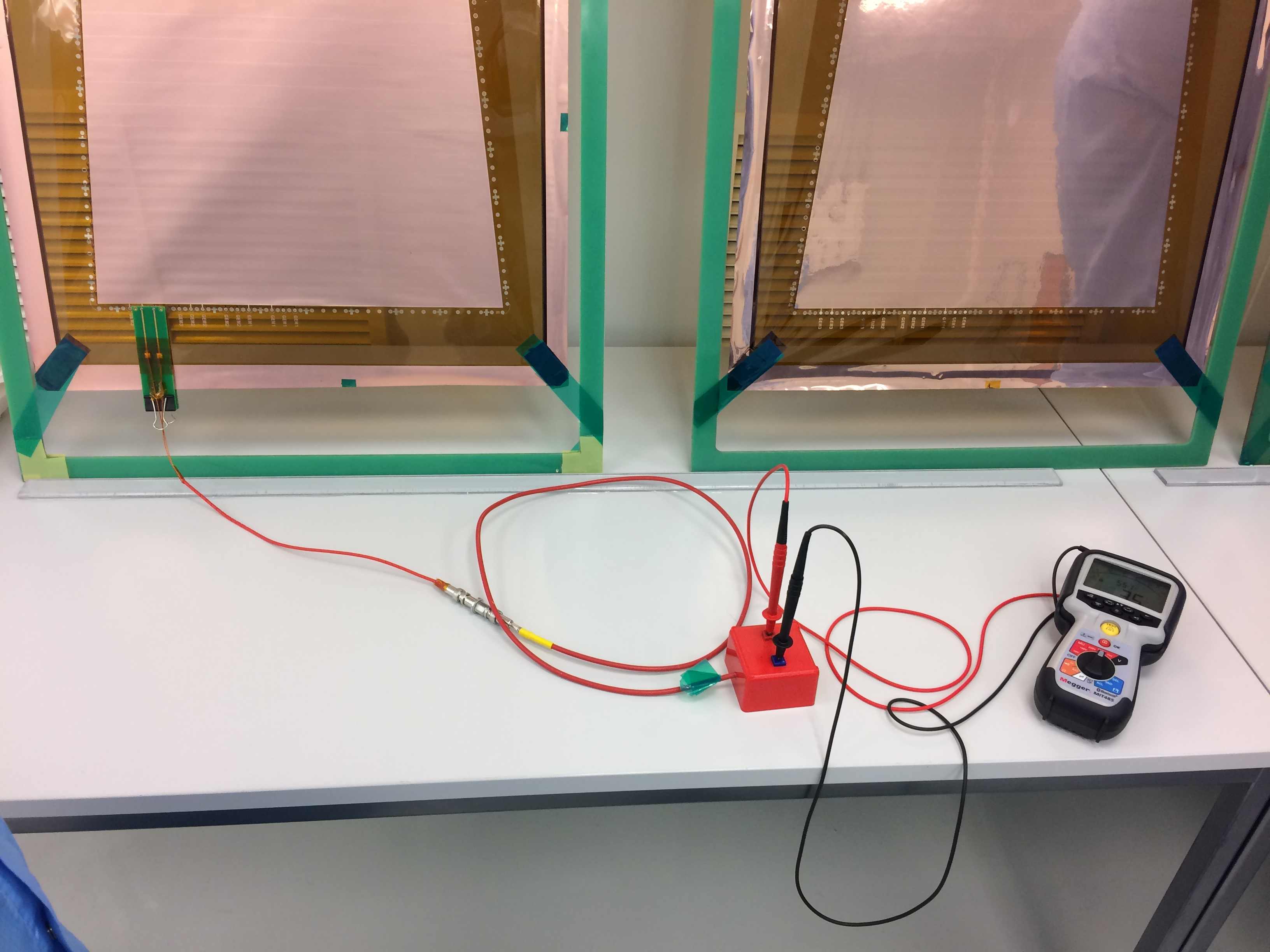} 
\caption{(left) GEM foil cleaning using adhesive roller. (right) Leakage current  measurement using a MIT 420 Megger.} 
\label{fig:adhesive_leakage}
\end{figure}

Before placing a GEM foil on the stack, it is first cleaned  using an antistatic adhesive  roller, which removes dust particles at the micron level  by its strong  sticking capacity as shown in Figure~\ref{fig:adhesive_leakage}~(left). This is followed by a leakage current measurement in which a potential difference of 550~V is applied across the foil using an insulation tester as shown in Figure~\ref{fig:adhesive_leakage}~(right). The applied potential produces a  very high electric field, typically of the order of 70-100~kV/cm, within the GEM holes and may lead to initial sparks due to the burning of any dust particles within the GEM holes. As such, this test may effectively help to clean the foils from leftover  impurities. In an environment with 30$\%$ of relative humidity (RH) or less, the maximum allowed value of the leakage current when a potential difference of 550V is applied across the foils is 35~nA. However, above 30$\%$ of RH, the leakage current can drastically rise above that level.  During this test, electrostatic charge is accumulated inside the GEM holes making them more likely to catch any remaining dust particles. Moreover, during the mounting of the stack, the GEM  foils are not  yet perfectly stretched, and if two adjacent foils in the stack come in contact, the whole energy accumulated on the foils can be released in a single point which may lead a damaged GEM foil. For this reason, at the end of the leakage current test, the GEM foils must be fully discharged by shorting the top  and bottom  electrodes. 

\begin{figure}[hbtp] 
\centering
\includegraphics[width=8cm, height=5cm]{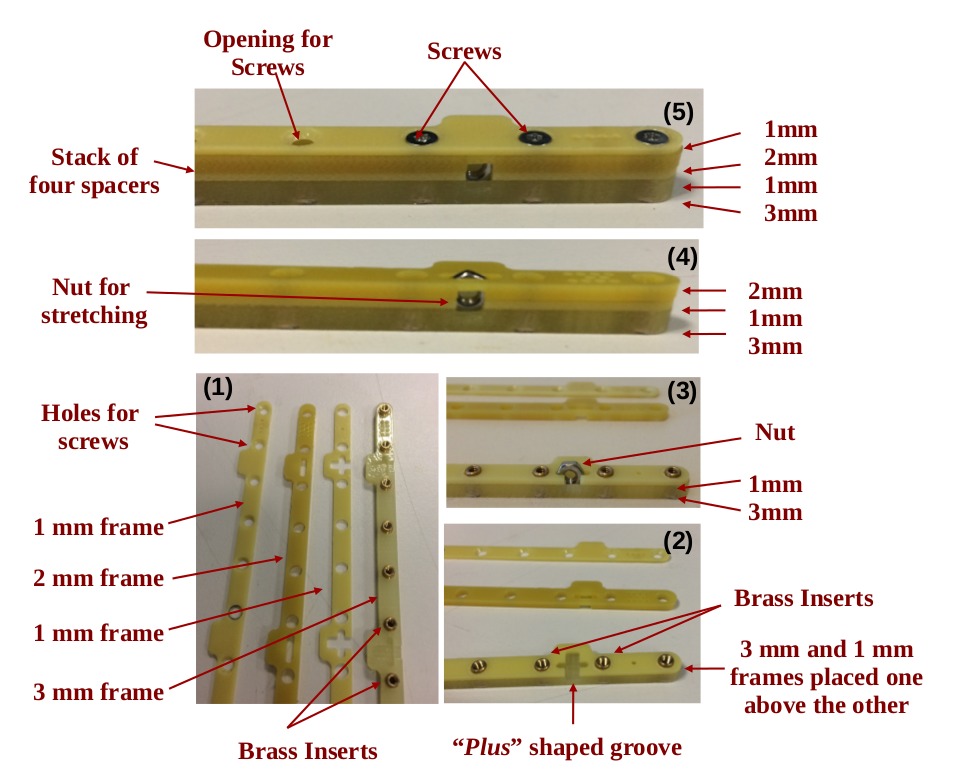}
\caption{(1) Shapes and mechanical structure of 4 different frame pieces, (2)  3 mm and 1 mm frame pieces are combined together with "+" shaped grove reserved for stretching nut, (3) placement of nut after stacking two frame pieces, (4) 3 mm, 1 mm, and 2 mm frame pieces combined in a specific order and (5) different frame pieces stacked together to maintain the gap configuration of 3/1/2/1 mm as demanded by GE1/1 geometry, similar frame pieces are combined to form four layers around the periphery of the GEM stack as depicted in Figure~\ref{fig:Internal_frames}.} 
\label{fig:Inside_frames}
\end{figure}

\begin{figure}[hbtp]
\centering  
\includegraphics[width=4.5cm, height=3cm]{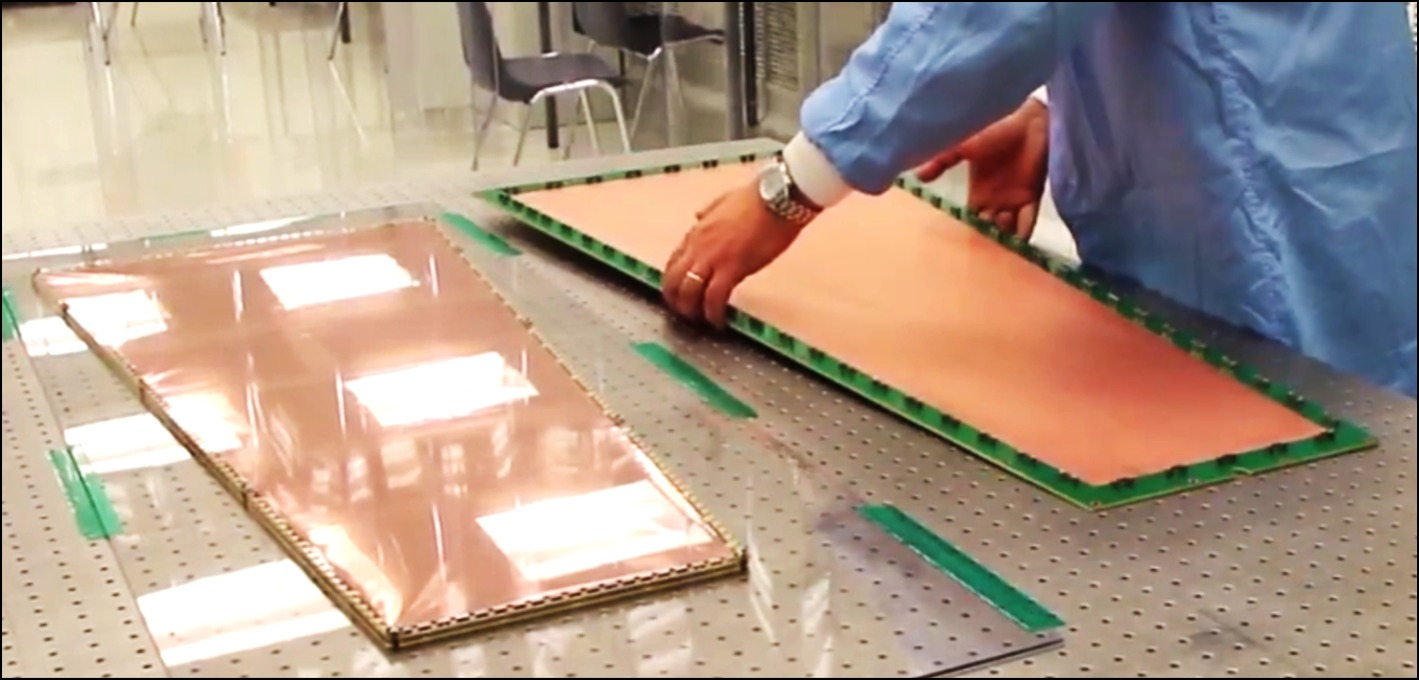}
\includegraphics[width=4cm, height=2.7cm]{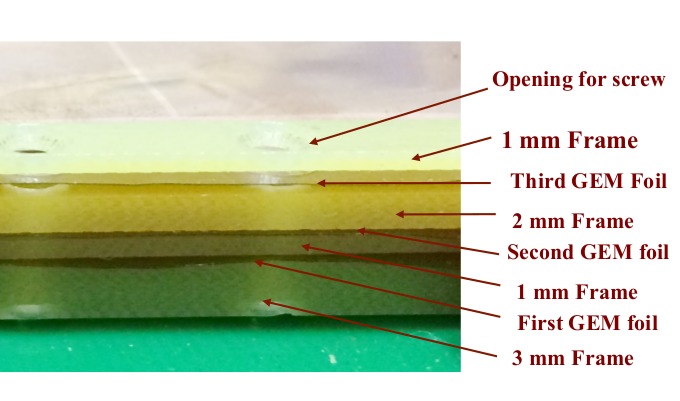}
\caption{(left) A GEM stack with a drift board and (right) the GEM stack structure from bottom to top.} 
\label{fig:stack}
\end{figure}

As was shown already in Figure~\ref{fig:GEM_FOil}, the edges of the GEM foils have a pattern of holes, which are used to attach the foils to  internal frames. The stack of three GEM foils is formed by sandwiching foils at their edges between the four layers of internal frames described in Section~\ref{Section_Internal_frames_design}. A detailed view of these  frames is shown in Figure~\ref{fig:Inside_frames}. The stack is held together by numerous small M 2 $\times$ 6 non-magnetic stainless steel screws, spaced about every centimeter along the frame and penetrating all the frame layers and foils. These screws   are tightened against small threaded M2 brass inserts. After the second GEM foil is placed on the  stack followed by the 2~mm frame, stainless steel nuts are embedded into the frames (at specifically designed "$+$" shaped grooves) every few centimeters around the periphery of the stack, with the axes of their threaded holes oriented parallel to the
plane of the inner frame. These nuts together with the pull-outs mounted on the drift board  and the corresponding screws are used in the stretching mechanism as will be  explained in  Section~\ref{stretching_Mechanism}. The GEM stack consisting of three foils with the 3$\slash$1$\slash$2$\slash$1~mm frames is finished by removing the dead area surrounding the active area  of the foils as shown in Figure~\ref{fig:stack}. 
%

\subsection{GEM foil stretching mechanism} 
\label{stretching_Mechanism}

\begin{figure}[hbtp]
\centering
\includegraphics[width=6cm, height=2.5cm]{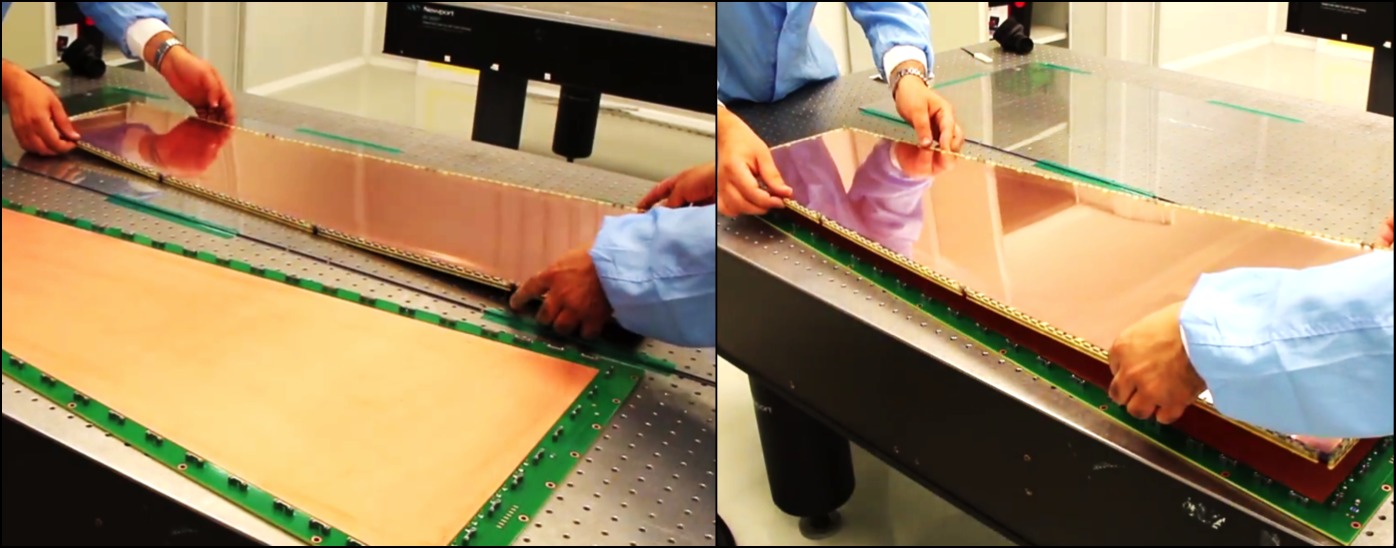}
\includegraphics[width=6cm, height=2.5cm]{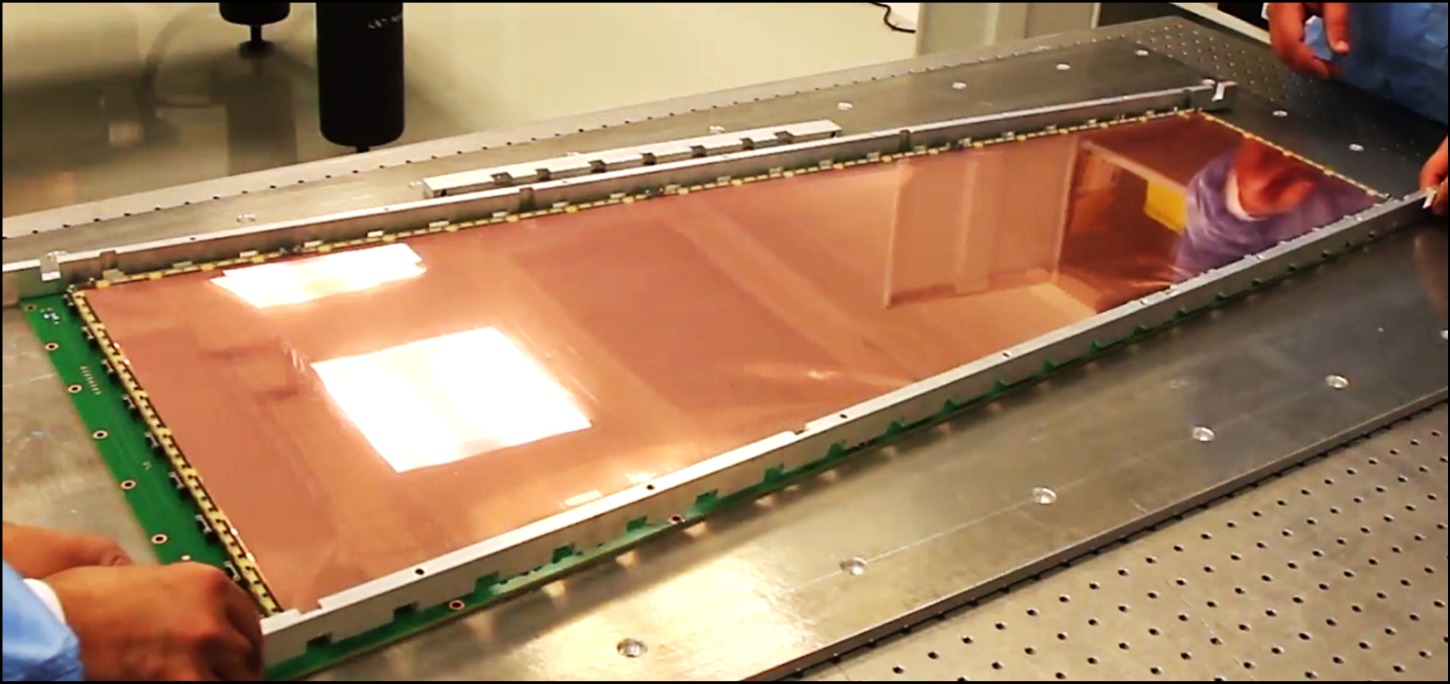}
\caption{(top) Placing the GEM stack onto the drift board. (bottom) The GEM stack on the drift board and fixation of the drift board with the assembly  jig.} 
\label{fig:Section_GEM}
\end{figure}

After the preparation of the three-layered GEM stack as described in Section~\ref{stack}, this stack is then  placed over the drift board as shown in Figure~\ref{fig:Section_GEM}~(top). The drift board is fixed against a special assembly jig made of aluminium bars which are  tightened with the help of fixation bolts as shown in  Figure~\ref{fig:Section_GEM}~(bottom).  The jig is used to keep the drift board flat by preventing deformations during the stretching of the GEM stack or during the fixation of the readout board against it when closing the chamber at the end of the assembly procedure.

\begin{figure}[hbtp]
\centering
\includegraphics[width=7cm, height=4.5cm]{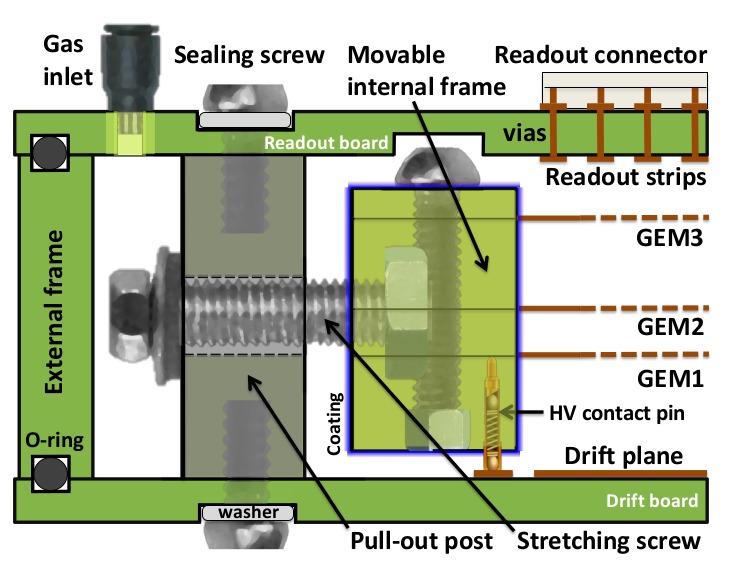}
\caption{Concept and mechanism employed to stretch the GEM foils in GE1/1 chambers~\cite{one_01}.} 
\label{fig:mechanism}
\end{figure}

\begin{figure}[h]
\centering
\includegraphics[width=7cm, height=4.5cm]{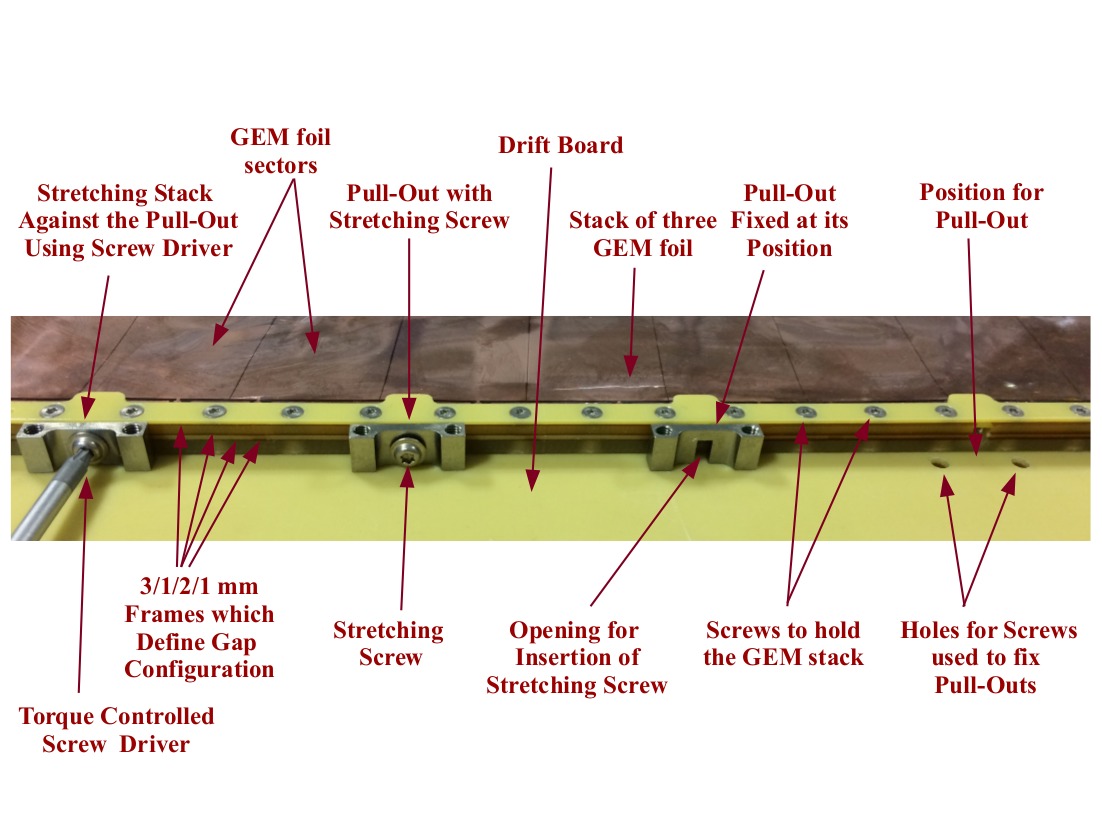}
\includegraphics[width=7cm, height=3cm]{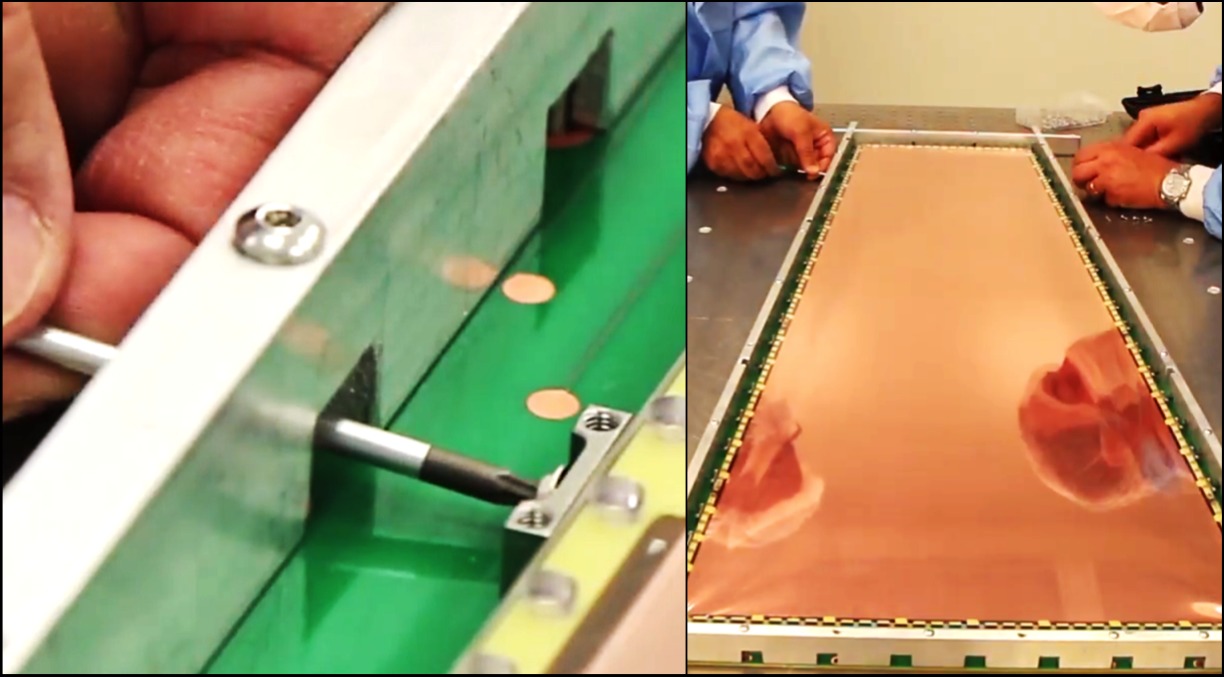}
\caption{(top) The implementation of the concept depicted in the Figure~\ref{fig:mechanism} by using various components to stretch the GEM
stack against the pull-outs. Stretching screws are allowed to pass through the stainless steel pull-outs and are recieved by nuts embeded into the inner
frame for tensioning the GEM foils in the stack. (bottom) The actual stretching is performed using a screw driver with an assembly jig mounted onto the drift board.} 
\label{fig:Stretching_Pic}
\end{figure}

Next, the entire GEM stack is mechanically stretched, i.e. all three foils at once, following the technique that is  conceptually visualized in  Figure~\ref{fig:mechanism}. The nuts inserted after the second GEM foil is put on the stack as explained in Section~\ref{stack} recieves M 2.5$\times$8/$\times$8 stainless steel screws that are inserted laterally into the  pull-outs located within the gas volume on the drift board as shown in Figure~\ref{fig:Stretching_Pic}~(top). 
The stack is then uniformly stretched against the pull-outs by amanually applying a controlled torque of about 8-10~cNm on the lateral screws, pulling the inner frame outwards towards the pull-outs as shown in Figure~\ref{fig:Stretching_Pic}~(bottom).  In order to  lower  the risk of forming "waves" in the GEM foils during the stretching of the stack, a specific order is followed in the tightening of the screws. Pairs of corresponding screws on opposite sides of the chamber are tightened simultaneously by two people.

Tolerances inherent in the stretching mechanism of the GEM foils and their  relative positioning can have an impact on the uniformity of the gas gain and
the detector timing  response. It is therefore crucial to ensure uniform stretching during assembly in order to achieve a uniform response across the detector surface. This uniform stretching is ensured by setting specifications on the torques applied to the pull-out screws during assembly. 
To determine and validate the optimal mechanical  tension applied to the GEM foils, a procedure involving Fiber Bragg Grating (FBG) sensors was used~\cite{three_03}. 

\subsection {Closing the chamber} 
After the foil stretching, a connectivity test between the gaps and across the foils is performed by measuring the impedance using a Mega Ohm Insulation tester. This is done by applying a potential difference of 550~V across the GEM foils and in between different gaps through the HV traces outside of the gas volume on the drift board. An impedance of more than 100~G$\Omega$ and 15~G$\Omega$ is expected between the gaps and across each GEM foil, respectively, for RH of 30$\%$ or less. The impedance may have lower values at higher values of humidity. 

\begin{figure}[hbtp]
\centering
\includegraphics[width=7cm, height=5.5cm]{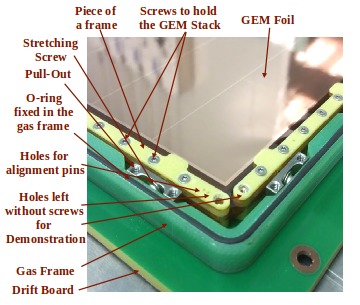}
\caption{Close-up view of a section of a GE1/1 detector with the GEM foil stack tensioned against the pull-outs mounted onto the drift board and surrounded by the outer frame equipped with O-rings in its grooves. The active chamber volume is ready to be closed with the readout board.} 
\label{fig:Close_Up_Section_GEM} 
\end{figure}

\begin{figure}[hbtp]
\centering
\includegraphics[width=5cm, height=3.7cm]{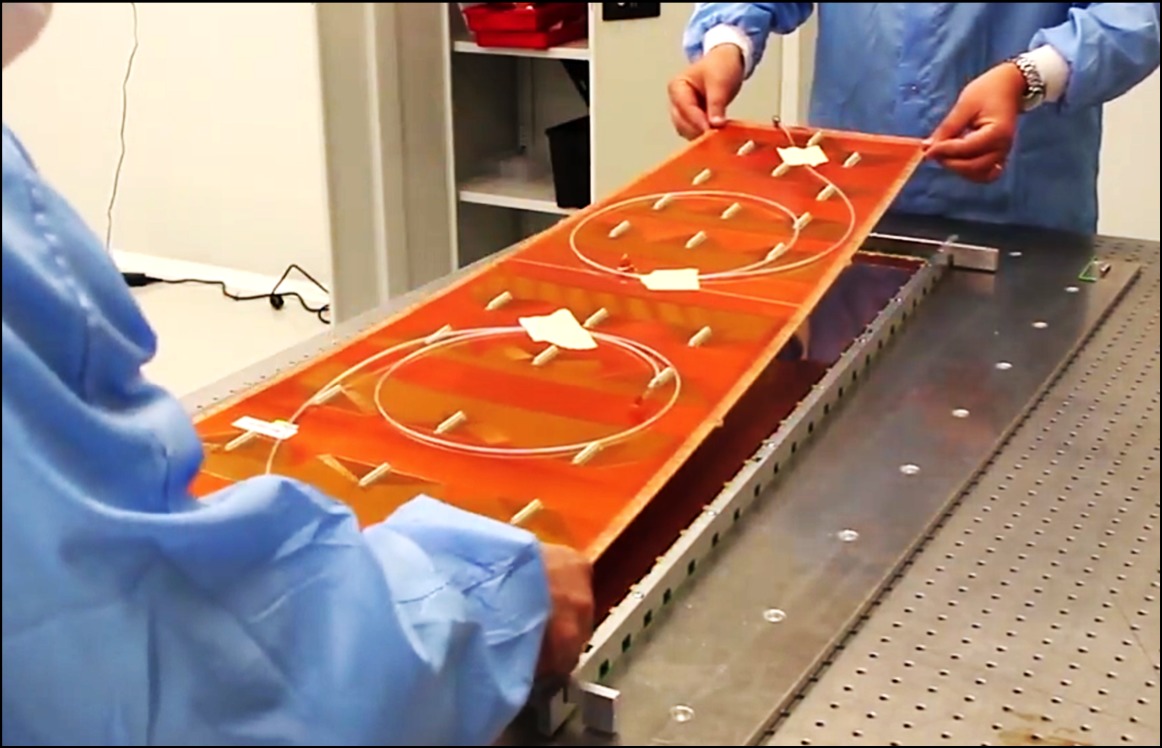}
\includegraphics[width=2.5cm, height=3.7cm]{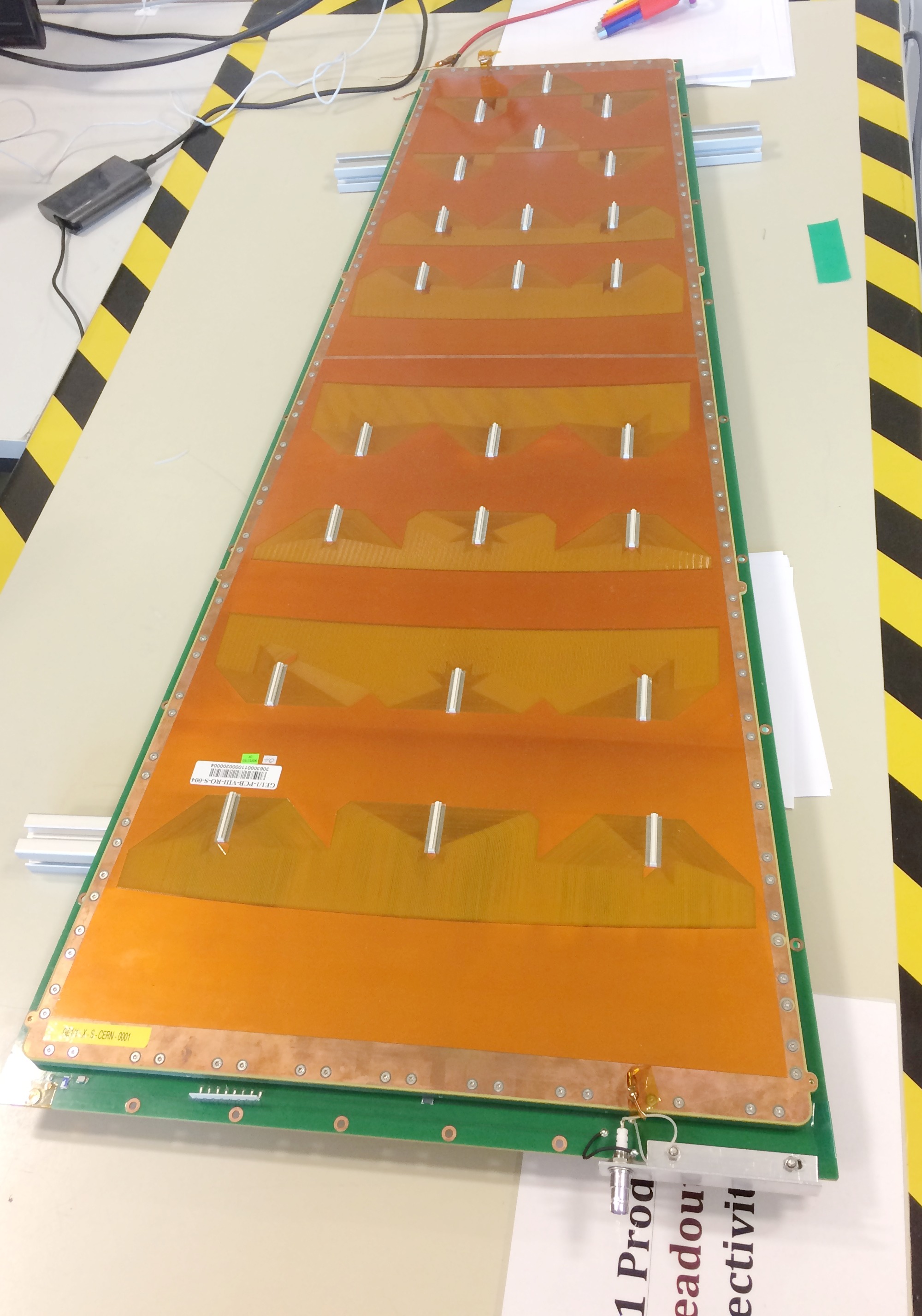} 
\caption{(left) Closing the chamber with the readout board. (right) A completed, closed GE1$\slash$1 chamber. } 
\label{LatestVersion}
\end{figure}

At the end of the assembly procedure, the external frame is placed around the tensioned GEM stack. It defines the limits of the gas volume as shown in
Figure~\ref{fig:Close_Up_Section_GEM}. The anode readout board is placed on top of this outer frame and is attached to the stainless steel pull-outs with
A2 stainless steel M3 $\times$ 6$\slash$ $\times$ 8 screws which are sealed with polyamide washers against the readout board as shown in Figure~\ref{LatestVersion}. This sandwiches the outer frame tightly between the drift and readout boards and provides a solid gas barrier. The final detector is shown in Figure~\ref{LatestVersion}.  

\section{Summary and outlook}
As part of its High Luminosity LHC upgrade project, the CMS Collaboration recently   approved the use of GEM technology to extend its muon system in the forward
region. In a first step, 144 new triple-GEM chambers (GE1/1) will be added to the  first muon endcap stations during the upcoming second LHC Long
Shutdown. After several years of R$\&$D that started in 2009, the design of these CMS GE1$\slash$1 chambers has been finalized.  The chambers are constructed using a novel mechanical stretching system requiring no gluing during assembly and no spacers inside the active area of the detector. This assembly technique reduces the chamber  construction time from several days to a few hours only.  Before reaching the final design, several full-size prototype chambers were produced using this  stretching technique and were successfully operated in test beams~\cite{one, two_01, four}. 

The final integration of the GE1/1 chambers into the CMS muon system is now being prepared. Following the design and assembly technique described here, ten chambers have been assembled, tested and installed inside the CMS experiment during the 2017 Extended Year End Technical Stop (EYETS) ~\cite{illariaIEEEpaper}. These chambers are providing the first operational experience with GEM
detectors inside CMS. The installation and integration time required later on for the full GE1/1 station can now be optimized.  The production and quality control of chambers for the full GE1/1 system is on schedule for installation in 2019.
\section*{Acknowledgements}
We gratefully acknowledge the support from FRS-FNRS (Belgium), FWO-Flanders (Belgium), BSF-MES (Bulgaria), BMBF (Germany), DAE (India), DST (India), INFN (Italy), NRF (Korea), LAS (Lithuania), QNRF (Qatar), DOE (USA) and the RD51 collaboration.


\clearpage 
\begin{thebibliography}{10}
\expandafter\ifx\csname url\endcsname\relax
  \def\url#1{\texttt{#1}}\fi
\expandafter\ifx\csname urlprefix\endcsname\relax\def\urlprefix{URL }\fi
\bibitem{CMSDetector} CMS Collaboration, JINST, 3 (2008) S08004. \\
\url{http://iopscience.iop.org/article/10.1088/1748-0221/3/08/S08004}
\bibitem{CMSMuonperf} CMS Collaboration, JINST, 13 (2008) P06015. \\
\url{http://iopscience.iop.org/article/10.1088/1748-0221/13/06/P06015}
\bibitem{illariaIEEEpaper} D. Abbaneo et al., "Operational experience with the GEM detector assembly lines for the CMS forward muon upgrade", IEEE Transactions on Nuclear Science (2018) submitted.
\bibitem{one_01} CMS Collaboration, CMS TDR, CERN-LHCC-2015-012 (2015). \\
\url{https://cds.cern.ch/record/2021453?ln=en}
\bibitem{Sauli} F. Sauli et al., NIMA. A386 (1997) 531. \\
\url{https://inspirehep.net/record/457577?ln=fr}
\bibitem{one} D. Abbaneo et al., IEEE Nucl. Sci. Symp. Conf. (2012) N14-137. \\
\url{http://ieeexplore.ieee.org/document/6551293/}
\bibitem{two_01} D. Abbaneo et al., IEEE Nucl. Sci. Symp. Conf., (2011) 1806-1810. \\
\url{http://ieeexplore.ieee.org/document/6154688/} 
\bibitem{two_01a} S. D. Pinto et al., JINST 4 (2009) 12009. \\
\url{http://iopscience.iop.org/1748-0221/4/12/P12009}
\bibitem{one_02} D. Abbaneo, et al., IEEE Nucl. Sci. Symp. Conf., (2011) 1909. \\
\url{http://ieeexplore.ieee.org/stamp/stamp.jsp?arnumber=5874107}
\bibitem{one_03} M. Staib et al., CERN-LHCC-2015-012 (2011), RD51-Note-004.
\bibitem{two} D. Abbaneo et al., JINST 8 (2013) C12031. \\
\url{http://iopscience.iop.org/1748-0221/8/12/C12031}
\bibitem{three} M. Tytgat et al., IEEE Nucl. Sci. Symp. Conf. (2011) 1019. \\ 
\url{http://ieeexplore.ieee.org/document/6154312/}
\bibitem{three_03} D. Abbaneo et al., 4th Int. Conf. on MPGD 174 (2018) 03002. \\
\url{https://doi.org/10.1051/epjconf/201817403002}
\bibitem{four} D. Abbaneo et al., IEEE Nucl. Sci. Symp. Conf. (2014) 1-8. \\
\url{http://ieeexplore.ieee.org/document/7431249/}
\bibitem{five} D. Abbaneo et al., NIMA 718 (2013) 383-386. \\
\url{http://dx.doi.org/10.1016/j.nima.2012.10.058}
\end{thebibliography}
\end{document}